\documentclass[apjl]{emulateapj}
\usepackage{apjfonts}
\usepackage{epsfig}

\def\beq{\begin{equation}}
\def\eeq{\end{equation}}

\begin{document}

\title{Statistical Description of a Magnetized Corona above a 
Turbulent Accretion Disk}

\author{by Dmitri A.\ Uzdensky \& Jeremy Goodman}
\affil{Princeton University}
\affil{Peyton Hall, Princeton, NJ 08544}
\email{uzdensky@astro.princeton.edu, jeremy@astro.princeton.edu}

\date{March 3, 2008}


\vskip 30 pt

\begin{abstract}
  We present a physics-based statistical theory of a force-free
  magnetic field in the corona above a turbulent accretion disk. The
  field is represented by a statistical ensemble of loops tied to the
  disk. Each loop evolves under several physical processes: Keplerian
  shear, turbulent random walk of the disk footpoints, and
  reconnection with other loops. To build a statistical description,
  we introduce the distribution function of loops over their sizes and
  construct a kinetic equation that governs its evolution.  This loop
  kinetic equation is formally analogous to Boltzmann's kinetic
  equation, with loop-loop reconnection described by a binary
  collision integral. A dimensionless parameter is introduced to scale
  the (unknown) overall rate of reconnection relative to Keplerian
  shear.  After solving for the loop distribution function numerically, 
  we calculate self-consistently the distribution of the mean magnetic
  pressure and dissipation rate with height, and the equilibrium shapes 
  of loops of different sizes. We also compute the energy and torque 
  associated with a given loop, as well as the total magnetic energy 
  and torque in the corona.  We explore the dependence of these quantities 
  on the reconnection parameter and find that they can be greatly enhanced 
  if reconnection between loops is suppressed.

\end{abstract}

\maketitle


\newpage
\section{Introduction}
\label{sec-intro}




Power-law components in the X-ray spectra of accreting black holes 
are attributed to hot, tenuous, quasi-spherical plasmas called {\it 
accretion disk coronae} (hereafter ADCe) by analogy with the solar
corona, though comptonization rather than atomic lines or bremsstrahlung 
is thought to be the main emission mechanism in accreting systems 
\citep{Bisnovatyi-Kogan_Blinnikov76, Liang_Price77}.
The vertical extent of ADCe is open to doubt since they are spatially 
unresolved and since the electron temperature inferred from high-energy 
spectral cutoffs, $T_e\sim 100\,\mbox{keV}$, is typically small compared 
to the virial temperature of the ions, $T_i\sim 100\,\mbox{MeV}$.  
Direct simulations of magnetorotational (hereafter MRI) turbulence in
radiation-pressure-dominated disks show very large density contrasts
and bulk velocities \citep{Turner_Stone_Sano02}, so perhaps the power-laws
are made within the disk itself \citep{Socrates_Davis_Blaes04}.

If a comptonizing corona indeed exists, then it is difficult to 
avoid the conclusion that it should be magnetically dominated.  
The electrons themselves cannot store much energy
\citep{Merloni_Fabian01}: their Compton cooling time is at most
comparable to the local dynamical time at luminosities $\gtrsim
10^{-2}L_{\rm Edd}$ and radii $\lesssim 20 GM_{\rm bh}/c^2$.  
Ions at virial temperatures would store much more energy but 
could not transfer it efficiently to the electrons by Coulomb 
collisions \citep{Rees_Begelman_Blandford_Phinney82}.  Also, 
as in the solar case, magnetic fields are probably needed to 
convert mechanical energy of the optically thick regions into 
coronal heat.  And since accretion is believed to be driven by 
MRI turbulence within the disk proper, it is expected that fields 
float up into the corona by Parker and interchange 
instabilities~\citep{Galeev_Rosner_Vaiana79,Tout_Pringle92}.

Despite widespread recognition of these points, modelers of ADC
emission rarely concern themselves with the dynamics of coronal
magnetic fields.  This is perceived to be too hard; certainly 
the long and arduous struggle to understand the heating of solar 
corona---based on much more abundant data---tends to discourage 
hopes of solving the corresponding problem for accreting systems 
any time soon \citep[\emph{e.g.}][]{Walsh_Ireland03,Klimchuk06}.  
Attempts to model coronae through direct three-dimensional (3D)
magnetohydrodynamic shearing-box \citep{Miller_Stone00,Hirose_Krolik_Stone06} 
and global \citep{Machida_etal00} simulations have been made, however.
To date, such simulations suggest that while a magnetically dominated 
region does form, its vertical extent, when defined by distributions 
of shear stress or dissipation rate, exceeds that of the optically 
thick regions only modestly.

There are nevertheless good reasons to expect dynamically dominant
coronae and to question the contrary evidence from simulations.  On
the one hand, a strongly magnetized corona may be required to extract
power from black hole spin \citep{Blandford_Znajek77} or from plunging
gas inside the marginally stable orbit \citep{Gammie99,Krolik99}, or
to drive a wind from the disk \citep{Blandford_Payne82}.  A less
widespread motivation, which however we feel strongly, is to reduce
self-gravity in the outer parts of accretion disks in active galactic
nuclei (at $r\gtrsim 10^3 GM_{\rm bh}/c^2$): a magnetized corona or
wind might transport angular momentum more quickly than stresses
limited to the optically thick layer, and thereby reduce the mass
density within the disk for a given accretion rate \citep{Goodman03}.

On the other hand, codes designed for pressure-dominated plasmas 
may not be reliable when applied to magnetically dominated coronae.
Shearing-box calculations cannot be trusted to represent magnetic
structures much larger than a disk scale height ($H$) unless 
\emph{all three} dimensions of the box are $\gg H$, which has not 
yet been achieved.  Global simulations are unable to resolve thin 
disks and extended coronae unless the grid is made coarser in the 
corona than near the midplane, which increases numerical diffusion 
in the corona. This is not serious when $H\sim r$, as for the innermost 
parts of near-Eddington or radiatively inefficient accretion disks, 
where the X-ray evidence for coronae is strongest; but it would be 
a severe limitation for simulations of the marginally self-gravitating 
parts of AGN disks, where $H/r\lesssim 10^{-2}$.  These problems of
spatial dynamic range will eventually be overcome with sufficient
computer power.

A more fundamental difficulty for simulations has to do with magnetic
reconnection.  We will argue that the efficiency of
reconnection\footnote{It would be natural to write ``rate of reconnection''
here, but that this phrase is often used to mean the speed at which 
field lines of opposing polarities approach one another in a localized 
reconnection event, whereas we are concerned here with the global 
efficiency of reconnection in reducing magnetic energy.} is crucial 
to the storage and dissipation of magnetic energy in ADCe, as it 
appears to be in the solar corona \citep{Parker83,Parker88,Klimchuk06}.  
Reconnection in solar flares is observed to be ``fast,'' meaning that 
annihilating field lines come together at a significant fraction of the 
local Alfv\'en speed.  This is not well understood; because of the very 
high conductivity of the corona, MHD predicts reconnection rates slower 
by many orders of magnitude \citep{Sweet58,Parker57}.  Collisionless
plasma effects outside conventional resistive MHD may be necessary
\citep[and references therein]{Uzdensky07a,Uzdensky07b}.  Another
mystery is what triggers fast reconnection, which does not always
occur immediately but seems to require special conditions that have
not been fully identified \citep{Klimchuk06}.  The overall rate of 
magnetic dissipation depends both on the rate at which reconnection 
events are triggered and on the speed of reconnection during such 
events.

These complexities are elided by the astrophysical MHD codes used 
to study magnetorotational turbulence.  Often in these codes,
reconnection is purely numerical, that is, produced by truncation
errors due to limited grid resolution.  Explicit diffusivities are
sometimes used, but if these are large enough make truncation errors
unimportant, then they are necessarily orders of magnitude larger 
than astrophysical reality.  Until recently, the general view among 
MRI simulators seems to have been that the microphysics of reconnection 
is unimportant.  This view may be inspired by analogies with the 
dissipation of kinetic energy in hydrodynamic turbulence, both 
compressible and incompressible.  Supersonic turbulence involves
shocks, and as is well known, shock dissipation is independent of
transport coefficients in the limit that these are small.  On the
other hand, three-dimensional incompressible turbulence involves
inertial cascades such that dissipation, though occurring on small
(viscous) scales, is entirely controlled by the dynamics on large
scales and proceeds at rates that are again asymptotically independent
of transport coefficients.  

Whether or when turbulent \emph{magnetic} dissipation is similarly
independent of small---and therefore numerically unresolved---scales
is an open question.  Recent work indicates that magnetorotational
turbulence is sensitive to the magnetic Prandtl number
$P_m\equiv\nu/\eta$
\citep{Lesur_Longaretti07,Fromang_Papaloizou_etal07}, at least in 
the range of fluid and magnetic Reynolds numbers ($Re\equiv VL/\nu$,
$Re_m\equiv P_m Re$) accessible to direct simulations.  It may be that
$Re_m$ and $Re$ become unimportant when both are sufficiently large,
but this has not yet been established even for kinematic dynamos,
where the field is dynamically unimportant on all scales by
construction \citep[and references
therein]{Boldyrev_Cattaneo04,Schekochihin_etal07}, much less for MRI
turbulence.  Even if this is true of systems in which fluid motions
dominate the energy density, as they are presumed to do near the
midplane of an accretion disk, the answer could be different for
magnetically dominated systems such as ADCe. Perhaps relevant here 
is the case of turbulence in the presence of a dominant mean field, 
which has become somewhat better understood since the seminal paper 
of \citet[hereafter GS95]{Goldreich_Sridhar95}.  Cascades do exist 
in such turbulence, with different scalings along and perpendicular 
to the mean field, so that the large scales are insulated from details 
at the resistive and viscous scales.  However, there are at least two
important differences between such systems and the nearly force-free
ADCe contemplated in this paper.  In the former systems, the
Alfv\'enic propagation time along the field is assumed to be longer
than the timescale of the cascade, so that the turbulent dynamics are
local; by contrast, communication along the entire length of a line-tied 
coronal loop is effectively instantaneous compared to the timescale 
on which energy is injected into the loop by footpoint motions.  
The magnetic dynamics are therefore nonlocal, and not appropriately 
characterized as turbulent; a better description is a progression 
of force-free equilibria driven by changes at the boundary.  
[See, however, \cite{Rappazzo_etal07} for a contrary view.]
Secondly, in GS95's Alfv\'enic cascades, the deviations from the
mean field are small, and the cascade does not affect the energy 
in the large-scale mean field.
In this paper, by contrast, we are concerned with coronal flux loops 
that reconnect with one another at large angles between their field 
lines. A significant fraction of the loops' magnetic energy may be 
liberated in such reconnection events, or in the relaxation to a new 
force-free equilibrium following topological changes brought about by 
reconnection.  

It might be hoped that the MRI simulations could predict the total
dissipation rate of the corona, if not its vertical distribution,
because the rate of injection of energy to the corona is determined 
at its base, where thermal and kinetic energies dominate. This is a 
false hope, however. The rate of work done on the corona by the disk 
is proportional to the magnetic stress tensor at its base, specifically 
the $rz$ and $\phi z$ components of the stress. Insofar as the corona 
is approximately force-free, its total energy is expressible in terms 
of a boundary integral involving the same stress components. Thus if 
the coronal energy and magnetic configuration are incorrectly calculated, 
then the coronal dissipation rate is probably also incorrect.
More concretely, because most of the kinetic energy available from 
the disk is the large-scale differential rotation rather than local 
turbulence, coronal field loops with large radial separations between 
their footpoints may be particularly important for the energy input; 
such loops are not possible in shearing-box simulations whose radial 
dimensions are no larger than the disk scale height, and even if the 
dimensions were increased, spurious reconnection might suppress the 
large loops.  In short, for geometrical reasons and because they do 
not model reconnection correctly, present-day MRI simulations may 
underestimate the energies and dissipation rates of disk coronae.

The dynamics of ADCe may not be fully understood without great progress 
on all of the fronts described above: more powerful computations, better 
understanding of fast reconnection, and of course more incisive observations.  
Since all of this may take years or decades to accomplish, our purpose here 
is to try to imagine, in a disciplined way, some aspects of that ultimate 
understanding.
Our approach is clearly indebted to \citet[henceforth TP96]{Tout_Pringle96}
but is richer in physical elements.  We model the ADC as an assembly
of closed magnetic loops with footpoints on the disk; open field is
probably important but is deferred to a later paper because we do not
wish to address winds here.  We do not resolve the dynamics within the
optically thick disk at all but treat the base of the corona as a
dynamic boundary.  It is assumed that the disk thickness is much less
than its radius in the parts of the ADC that we model and that the
loop lengths ($L$) lie at intermediate scales ($H\ll L\ll r$).
Thus the lower boundary is conceived as an infinite plane.  
Small new loops are injected from this boundary, and existing loops are
energized by the Keplerian shear if their footpoints lie at different
radii.  Loops reconnect in pairs according to prescribed rules with a
frequency scaled relative to the shear by a dimensionless parameter.  
Each reconnection results in a new pair involving the same four footpoints
differently connected, rather than a single loop as in~TP96.  The loop
population is described by a distribution function over the length and
orientation of the displacement from the negative to the positive-polarity 
footpoint (rather than the length alone as in TP96), and the processes 
of injection, stretching, and reconnection are incorporated in an 
integro-differential kinetic equation for the evolution of the loop 
distribution function.

Our model is not appropriate for all forms of ADCe.  Following
\citet{Liang_Price77}, we presuppose a ``sandwich'' geometry in which
the optically thick but geometrically thin disk coexists with its
corona at the same radius.  Such a geometry seems most appropriate to
high/soft and very high states of galactic X-ray sources where the
X-ray continuum shows both thermal and power-law features, and also to
cases, both galactic and extragalactic, where a relativistically broad
iron K$\alpha$ line indicates that an optically thick disk, fluorescing
under illumination by hard X-rays from the corona, extends down to the 
marginally stable orbit or beyond
\citep[\emph{e.g.}][]{Wilms_Reynolds_etal01,Miller_Fabian_etal02}.  
In low/hard states where only a power-law is seen, it may be that 
the inner parts of the thin disk have evaporated so that those regions 
are all ``corona'' \citep{Esin_McClintock_Narayan97}; alternatively,
the thin disk may persist down to the marginally stable orbit, but
the corona may take the form of a mildly relativistic outflowing
wind or jet, whose emission is directed away from the disk
\citep{Beloborodov99,Miller_Homan_etal06}.  The coronal model presented here
would not apply to either of these cases without substantial modification.

In this paper, we explore what controls the magnetic energy,
integrated stress, and dissipation rate of the corona; specifically,
we explore the roles of shear and reconnection in this balance.  Our
overarching motivation is to determine under what conditions the
corona contributes importantly to outward angular momentum transport:
that is, to the torque that drives accretion through the disk.
Secondary goals are to examine the distribution of energy, stress, and
dissipation with height.  We recognize that because of a number of
questionable assumptions and simplifications, our model will hardly be
the last word on this subject.  We hope , however, at least to set up 
a target for future simulations to aim at, and to draw attention to
certain quantities that could be extracted from existing simulations,
such as the rate of emergence of flux dipoles (small loops) from the
disk.

The outline of our paper is as follows. \S~\ref{sec-statistical-model} 
introduces our conception of magnetic loops and the loop distribution 
function.  \S~\ref{sec-self-consistent} explores some properties of 
loops in equilibrium with the mean-field pressure exerted by neighboring 
loops: shape, maximum height, and energy.  \S~\ref{sec-kinetic-eqn}
constructs the kinetic equation, including the rules for reconnection
and other processes important to the evolution of the loop distribution.  
The numerical set-up used to solve the loop kinetic equation, including 
the boundary conditions at both small and large scales, is described 
in~\S~\ref{sec-numerical}. In the same section we also present numerical 
solutions and discuss their implications for ADCe.
In~\S~\ref{sec-discussion} we discuss limitations of our model in the light 
of these results, and we indicate ways in which the model might be made 
more realistic.  We attempt to relate what we have done to the present 
theoretical understanding of ADCe, and we discuss how simulations in 
the near future might be used to calibrate some features of our model, 
for example the rate of emergence of small loops from the disk.  
Finally, \S~\ref{sec-conclusions} summarizes our main conclusions.


\section{Statistical Description of the ADC Magnetic Field}
\label{sec-statistical-model}


As noted in \S1, there are fundamental physical similarities between 
the formation processes of the solar and accretion disk coronae.
At the same time, one has to keep in mind the important differences 
between them. These include: (1) differences in the underlying 
sub-photospheric turbulence (thermal convection vs.\ MRI); 
(2) a strong large-scale differential rotation and shear in 
disks, whereas small scales dominate the shear in the Sun;
(3) the possibility of strong magneto-centrifugally- and radiation-driven
winds from an accretion disk, as compared to the relatively weak thermally 
driven solar wind; (4) Compton cooling of the disk corona in black hole 
systems; (5) a greater separation of spatial scales in thin disks 
between the disk thickness and its radius, as opposed to the solar case, 
where the convection zone spans a significant fraction of the Sun's radius, 
which results in the generation and emergence of very large magnetic 
structures associated with sunspots in active regions. Finally, we have 
only one Sun, whereas there is a large variety of astrophysical 
accretion-disk systems, some of which have coronae. 

In addition to the above, there is an important basic difference in 
our observational capabilities: whereas the Sun is so close that we 
can spatially resolve {\it individual} events and structures in the 
solar corona (such as flares, loops, etc), such resolution is not available 
for ADCe. Therefore, we can study only the spatially-integrated spectral 
and timing properties of the disk corona. This fact provides a strong 
motivation for focusing on a {\it statistical description} for the 
magnetic field in the~ADC.

Here is, briefly, the basic physical picture of the magnetized corona 
above a turbulent accretion disk. The corona is a dynamic, self-organized 
system that can be represented by a statistical ensemble of flux loops 
\citep{Tout_Pringle96,Hughes_Paczuski_etal03}.
The loops continuously emerge out of (and submerge into) the disk as a result 
of magnetic buoyancy. Once above the surface, they constantly evolve due to 
a number of physical processes. 
They are twisted and stretched by the differential keplerian rotation and 
by the random motions of their footpoints on the disk's surface, which causes 
the individual loops to inflate. As a result, the magnetic field in the 
corona becomes non-potential and highly stressed; an appreciable amount 
of free magnetic energy can thus be stored in the corona. However, in 
the process of twisting and expansion the loops may undergo internal 
disruptions due to MHD instabilities and also may reconnect with other 
loops. Such relaxation events manifest themselves as flares; they bring 
the field closer to the potential state and thus enable the inflation 
process to resume. At the same time, reconnection between loops sometimes
produces more spatially-extended magnetic structures in the corona (the 
coronal ``inverse cascade''). Finally, all these complicated processes 
occur repeatedly over and over, simultaneously on various spatial scales.
Thus, the corona can be viewed as a {\it boiling magnetic foam}, in which 
magnetic loops repeatedly swell and grow because of the magnetic energy 
pumped into them by the footpoint motions and then snap and contract back 
due to reconnective disruptions and sometimes merge to form bigger structures.

What is the appropriate statistical language for describing a chaotic, 
highly intermittent magnetized corona above a turbulent accretion disk?
At a most general level, we describe the corona by an ensemble of some 
{\it elementary magnetic structures}. Each of these fundamental individual 
constituents of the corona is characterized by a small set of primary physical 
parameters. The individual elements evolve with time according to certain 
rules that reflect the relevant physical processes that we believe are most 
important in shaping the corona. These processes generally include various 
forms of interaction between elements. Mathematically, the evolution rules 
are represented by stochastic (Langevin) equations of motion of magnetic 
elements in the primary parameter space.  Since we are interested in a 
statistical description, we introduce a distribution function of our magnetic 
elements in the primary-parameter phase-space. Correspondingly, one of our 
main goals is to derive the equation for the evolution of this distribution 
function, using the equations of motion of individual elements. This is done 
by analogy with the way the Boltzmann kinetic equation for the particle 
distribution function in a gas is derived in Statistical Mechanics, but 
obviously is more {\it ad hoc} in our case. Finally, there are several 
important integral quantities in our theoretical framework, which are 
related to moments of the distribution function. These self-consistent 
quantities represent the mean-field interaction between the magnetic 
elements and they affect the evolution of the distribution function.

\begin{figure}[t]
\plotone{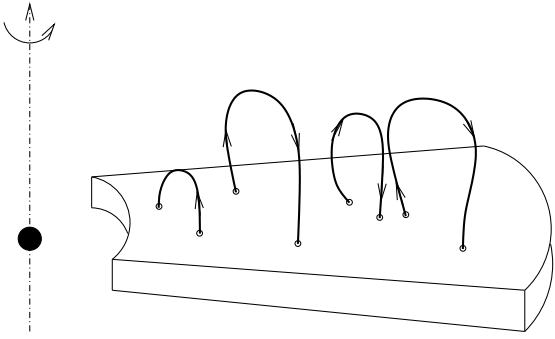}
\figcaption{Schematic view of ADC as an ensemble of many magnetic loops.
\label{fig-corona}}
\end{figure}


\subsection{The Loop Distribution Function}
\label{subsec-F}

Based on the above discussion, we shall now build a calculable model 
of the corona. Our first task is to select the most appropriate and
most fundamental elementary magnetic constituents of the corona.
We shall then need to select the most natural set of parameters
describing these elements.

\begin{figure}
\plotone{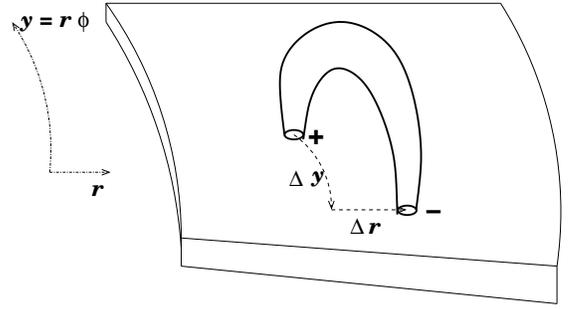}
\figcaption{Closed magnetic loop as the main structural element 
of a magnetized accretion disk corona.
\label{fig-loop-1}}
\end{figure}

Guided by the analogy with the solar corona, we shall use simple 
(anchored in the disk at both ends, see Fig.~\ref{fig-loop-1}) 
magnetic loops, or flux tubes, as our fundamental  magnetic elements 
--- the main structural constituents of the corona. This choice is 
influenced by the existing theoretical work in both solar physics 
\citep[\emph{e.g.}][]{Hughes_Paczuski_etal03}, and also in ADC 
\citep[]{Tout_Pringle96}. 
Such loops represent the closed magnetic field corresponding 
to zero net vertical flux through the disk. This is a natural 
assumption for the case when the magnetic field in the corona 
comes from the flux emergence of the field generated by the dynamo 
in the disk itself. In this paper we shall assume that this is indeed 
the case. In principle, however, one may also wish to consider a more 
general situation where, in addition to the closed coronal loops, there 
is also an external large-scale open magnetic field through the disk,
such as may be coming from the central star or the interstellar medium 
(see \S~\ref{subsec-open}).

For simplicity, we shall characterize each loop by only two primary parameters:
{\it (i)} the radial footpoint separation $\Delta r = r_{-}-r_{+}$,
and 
{\it (ii)} the azimuthal footpoint separation 
$r\Delta\phi = \Delta y =y_{-}-y_{+}$ (see Fig.~\ref{fig-loop-1}).
Thus, we shall measure $\Delta r$ and~$\Delta y$ from the ``+'' magnetic 
footpoint to the ``-'' magnetic footpoint. This means that $\Delta r$ 
and~$\Delta y$ can be positive or negative depending on the orientation of 
the loop. Alternatively, sometimes we will use an equivalent representation
in terms of the loop's projected length (the distance between the loop's 
footpoints): $L\equiv (\Delta r^2 + \Delta y^2)^{1/2}$; and the orientation
angle, $\theta$, measured clockwise with respect to the toroidal direction: 
$\theta \equiv \arctan(\Delta r/\Delta y)$.


In a more general description, one may enlarge the parameter 
space to include additional parameters, such as the magnetic 
flux~$\Delta\Psi$ contained within a loop, or the loop's twist
(see~\S~\ref{subsec-twisted}). 
However, since our goal here is to build the simplest version of this 
already very complicated theory, we shall assume that all the loops 
have the same magnetic flux~$\Delta\Psi$ and, furthermore, that they 
are not twisted. The latter assumption means that the magnetic field 
within each loop is purely potential, {\it i.e.}, that the bulk of the 
corona is nearly current-free and all the coronal currents flow along 
inter-loop boundaries.

In addition to the above two primary parameters, we shall also need 
some secondary parameters describing a given loop, such as the loop's
overall shape, its maximum height~$Z_{\rm top}$ above the disk; and 
its thickness at a given height, $d(z)$. 
These quantities will be useful for estimating the loop expansion rate and
for analyzing binary interaction (reconnection) of loops with each other.
In our model, these secondary parameters are uniquely determined by the 
primary ones in combination with the self-consistent mean field~$\bar{B}(z)$
(see \S~\ref{subsec-Bbar}).

Note that a loop carrying a finite flux $\Delta\Psi$ has a certain finite
thickness (in the radial and azimuthal directions) at the disk surface. 
For a typical loop, this thickness is  generally of order~$H$. Therefore, 
instead of a pair of footpoints that 
a field {\it line} would have, a finite-thickness {\it loop} 
has a pair of footspots. Thus, we need to be a little bit more 
precise in our definitions of~$\Delta r$ and~$\Delta y$. We shall 
define them as the radial and azimuthal separations between the 
centers of the two footspots, that is between the two footpoints 
of the central field line of the loop (magnetic axis for a twisted loop).

Following \cite{Tout_Pringle96}, we introduce the {\it loop distribution 
function}, $F(\Delta r,\Delta y)$, defined so that $F(\Delta r,\Delta y) 
d\Delta r d\Delta y$ is the number of loops with the values of primary 
parameters in the range $([\Delta r,\Delta r + d\Delta r],
[\Delta y,\Delta y + d\Delta y])$, per unit disk area. Alternatively,
we may write the distribution function in terms of loop length 
$L=(\Delta r^2 + \Delta y^2)^{1/2}$ and orientation 
$\theta=\arctan(\Delta r/\Delta y)$, {\it i.e.}, $F(L,\theta)$.
The overall normalization of the distribution function is determined
by the requirement that loops cover the entire disk surface; it will
be discussed in more detail in~\S~\ref{subsec-Bbar}.


\subsection{Role of Magnetic Reconnection in ADCe}
\label{subsec-recn-role}

At the most basic level, the corona (either solar or ADC) is perfectly 
conducting almost everywhere. However, as it evolves driven by the complex 
turbulent motions of the magnetic footpoints on the surface, the corona 
may develop numerous current sheets on a variety of scales \citep{Parker72, 
Parker83}.
These current sheets are possible sites of dissipation of magnetic energy 
{\it via} reconnection. In fact, reconnection is one of the most essential 
nontrivial physical processes that govern the complex dynamical behavior 
of the corona. In particular, it controls the vertical extent of the corona
({\it e.g.}, the coronal magnetic scale-height~$H_B$).
Indeed, if reconnection were too efficient, then the coronal field would 
be nearly potential and~$H_B\sim H$; then, the free magnetic energy stored 
in the corona would be small, as would the magnetic dissipation rate.

On the other hand, if no reconnection were allowed at all, then, 
magnetic loops would, over time, grow bigger and bigger in height 
because of the differential Keplerian rotation. Unable to dissipate, 
magnetic energy would continuously accumulate in the corona as the 
characteristic magnetic scale-height~$H_B$ increases, essentially 
linearly in time. This phase would continue until~$H_B$ becomes 
comparable to the disk radius, $H_B\sim r\gg H$. After that, radial 
gradients become important and the subsequent evolution  would enter 
a qualitatively different regime characterized by accelerated expansion 
of the magnetic loops, which effectively would become open, perhaps even 
in a finite time \citep{vanBallegooijen94,Lynden-Bell_Boily94,Aly95,
Sturrock_Antiochos_Roumeliotis95,Uzdensky02b}.
The corona would then consist of a dense forest of open flux tubes of 
alternating polarity separated by a multitude of current sheets. 
Unless there are significant mass-loaded winds (violating the force-free 
assumption), the power pumped into the corona would then go down, and the 
accumulated free magnetic energy would saturate at a value corresponding 
to a fully open (split-monopole) magnetic field~\citep{Aly91,Sturrock91}.
Although this asymptotic energy would be very large, of order~$r/H$ larger 
than that of the fully closed potential field, it would still remain finite;
this is because the toroidal magnetic field at the disk surface and hence 
the work done on the coronal magnetic field by the Keplerian disk shear 
would both go to zero.  Similarly, the angular momentum exchange between 
different parts of the disk due to coronal loops would also go down. As 
new flux tubes emerged from the disk, the magnetic forest would become ever 
more dense. The corona would thus look very different from what we expect. 
We thus see that reconnection is necessary for maintaining a {\it meaningful}
statistical steady state. It enables open field lines to close back and thus 
restores the magnetic connection between different parts of the disk. 
This, in turn, facilitates angular-momentum transport {\it via} the 
coronal magnetic field [coronal~MRI, \citep{Goodman03}; see also \citep
{Heyvaerts_Priest89,Pavlidou_etal01}]. Strong magnetic dissipation and 
large torque thus require some intermediate reconnection efficiency, 
neither so rapid as to keep the field nearly potential, nor so slow as 
to allow it to become fully open; in both limits, the torque and energy 
dissipation rate vanish.

Another reason why reconnection is important is that a growing magnetic
loop may reconnect with another one connected to a very different place 
on the disk. This process may lead to an {\it ``inverse cascade''} of 
magnetic loops \citep{Tout_Pringle96}. It is an important avenue towards 
building up a population of loops with large radial footpoint separation. 
Indeed, whereas Keplerian differential rotation increases the azimuthal 
footpoint separation of a loop, it does not affect its radial footpoint 
separation. Therefore, without reconnection, the radial footpoint separation 
of a coronal flux loop would change only relatively slowly by the random 
walk of its footpoints due to the underlying disk turbulence. In principle,
the footpoints will drift radially apart in direct response to the angular 
momentum transfer by the coronal loop itself (we call this process 
``coronal~MRI''). The characteristic velocity of this drift is on 
the order of $B^2/4\pi\Sigma \Omega$, where $\Sigma$ is the surface 
density of the disk. The resulting relative increase in~$\Delta r$ 
on the rotation-period timescale is of the order of $\delta\Delta r/\Delta r 
\sim B^2/4\pi\Sigma \Delta r \Omega^2$. Using $\Sigma\simeq H\rho$ and
$H\sim c_s/\Omega$, where $c_s$ and~$\rho$ are the sound speed and the 
gas density within the disk, we can estimate that $\delta\Delta r/\Delta r 
\sim (V_A/c_s)^2 \, H/\Delta r \equiv \beta^{-1}\, H/\Delta r$, where 
$V_A^2 \equiv B^2/4\pi\rho$ is the Alfv\'en speed within the disk.
Thus, since we are mostly interested in large loops, $\Delta r\gg H$,
we see that $\Delta r$ cannot grow appreciably without reconnection.
{\it Thus reconnection is necessary for the coronal ``inverse cascade''.}

In addition, magnetic reconnection in the corona regulates the fraction 
of the magnetic flux that is open at any given time and also the 
effective  radial transport of a large-scale vertical magnetic field 
\cite{Spruit_Uzdensky05,Fisk05}. Both of these processes are important 
for establishing large-scale disk outflows.

Finally, as in the solar corona, reconnection is believed to be the main 
mechanism of releasing the accumulated magnetic energy, leading to coronal
heating and observed high-energy coronal emission.


\section{The Self-Consistent Corona}
\label{sec-self-consistent}


In principle, the loop distribution function~$F(L,\theta)$ contains 
enough information to fully describe the statistical magnetic structure 
of the corona. This means that, once ~$F(L,\theta)$ is known, one should
be able to answer most of the questions posed in~\S\ref{sec-intro}.
In particular, one should be able to derive the actual shapes and heights,
$Z_{\rm top}(L)$, of coronal loops, the distribution of magnetic energy 
with height, $\bar{B}^2(z)/8\pi$, the energy~$\mathcal{E}(L)$ associated 
with a loop of a given size, the torque transmitted by the coronal magnetic 
field, etc. In this section we demonstrate how to do all this.


\subsection{Equilibrium Shape of a Loop in a Stratified Atmosphere}
\label{subsec-shape}

\begin{figure}[h]
\plotone{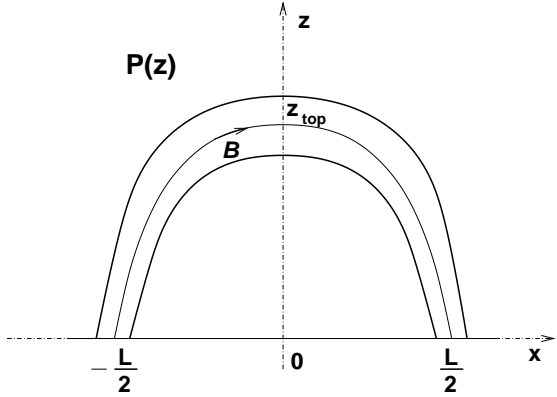}
\figcaption{Untwisted isolated loop confined by external isotropic
pressure~$P(z)$.
\label{fig-loop-2}}
\end{figure}

First, we shall work out the correct shape of an isolated slender 
(with a cross-sectional diameter $d\ll L$) untwisted loop~$\mathcal{A}$ 
carrying magnetic flux~$\Delta\Psi$, immersed in a medium with some isotropic 
but, in general, nonuniform pressure~$P(z)$ (see Fig.~\ref{fig-loop-2}).
This pressure represents the confining magnetic pressure of all other loops; 
thus, for actual calculations, it will be convenient to write~$P(z)$ as 
$P(z)\equiv\bar{B}^2(z)/8\pi$.
The shape of the loop is then determined by the requirement that the loop 
be in magnetostatic equilibrium with this external pressure.

First, the local pressure balance across the loop gives 
us the magnetic field strength inside it as a function of
height:
\beq
B(z)  = \sqrt{8\pi P(z)} \equiv \bar{B}(z)\, .
\label{eq-pressure-balance}
\eeq

Then, since the magnetic flux is constant along the loop,
we can write the local cross-sectional area~$a(z)$ of the
loop in terms of~$\bar{B}(z)$:
\beq
a(z) = {\Delta\Psi\over{\bar{B}(z)}} \, .
\eeq

Now let us discuss  equilibrium shape~$x(z)$ of a slender loop as 
a whole, represented by the shape of the loop's central field line. 
First, we would like to note that, for a curved but untwisted loop 
confined by a {\it uniform} external pressure~$P_0$, it is impossible 
to find an equilibrium shape.
Indeed, since the field inside the loop has no twist and is 
purely axial ({\it i.e.}, runs along the loop), it is a potential 
field produced by perpendicular (to the direction of the loop) 
currents flowing on the loop's surface. Therefore, at any given 
location along the loop, the magnetic field strength is slightly 
nonuniform in the cross-loop direction: it drops off as~$1/R$, 
where~$R$ is the local curvature radius (``major radius'' in 
tokamak terminology). In other words, the magnetic force balance 
between the magnetic pressure and the magnetic tension inside a 
curved loop means that there must be a magnetic pressure gradient 
to balance the tension force due to the curvature. Therefore,
the magnetic field on the underside (the ``inboard'', in tokamak 
terminology) of the loop is larger than that on the upside (the 
``outboard''). On the other hand, however, the magnetic field at 
each point on the surface of the loop has to be in pressure balance 
with the external pressure~$P(z)$. If~$P(z)=P_0={\rm const}$, this 
pressure balance means that the magnetic field has to be uniform 
along the loop's boundary. Thus, we have a clear incompatibility 
between the assumption that the loop is curved but untwisted and 
the condition of local force balance with an external uniform 
pressure.

This phenomenon can also be understood using the notion of Pfirsch--Schl\"uter 
currents, a well-known concept in tokamak physics. Since the magnetic field 
inside an untwisted flux tube is strictly in the axial direction (along the 
tube), the currents that produce it flow on the skin of the tube in the 
perpendicular direction. But since the currents have to close, and since 
the tube is curved, the surface current density ({\it i.e.}, the current 
per unit length along the tube) is bigger on the underside than on the upside. 
Then, according to Ampere's law, the magnetic field is also stronger 
on the underside, and thus cannot be in pressure equilibrium with a
uniform external pressure on both sides simultaneously. The only way
a curved loop can be in equilibrium with uniform external pressure, is
when the surface current density is also uniform. Since the current has
to be conserved, this requires that some surface current should also 
flow along the loop (the so-called Pfirsch--Schl\"uter current), and
hence the magnetic field must be twisted.


These considerations show that an untwisted flux tube confined by a uniform 
external pressure has to be straight. If, however, the pressure is not 
uniform, then an equilibrium shape for a curved tube can be found, as we 
now show.

\begin{figure}
\plotone{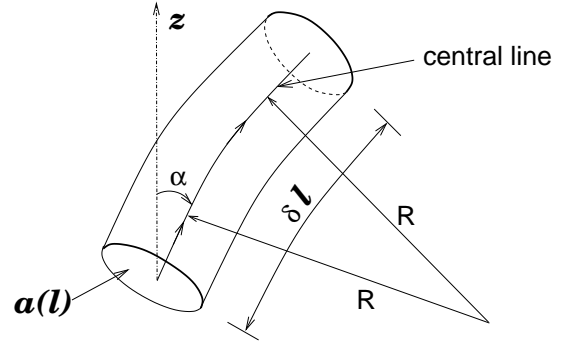}
\figcaption{Infinitesimal loop segment.
\label{fig-force-balance}}
\end{figure}

Let us consider a case with pressure~$P(z)$ decreasing monotonically 
with height. Consider a slender loop in the $(x,z)$ plane, symmetric 
with respect to $x=0$, with its two footpoints at $x=\pm\,L/2$ 
(see Fig.~\ref{fig-loop-2}). 
The shape of the loop as a whole, described by a symmetric function~$z(x)$, 
is determined by the perpendicular (to the magnetic field) force-balance 
between tension and the pressure-gradient forces. The force-balance is 
established at every point along the loop. To analyze it at a given
location on the loop, let us consider a small loop segment. We shall denote 
the arc-length along the loop, measured from its left footpoint, by~$l$.
For clarity of discussion, let us represent the loop segment locally 
by a slightly curved cylinder of length~$\delta l\ll L$, cross-sectional 
area~$a(l)$, and curvature radius~$R(l)$ (see Fig.~\ref{fig-force-balance}). 
Let us also introduce the angle~$\alpha(l)$ between the magnetic field and 
the vertical direction:
\beq
\cos\alpha = {dz\over{dl}} \, .
\label{eq-def-alpha}
\eeq
For definiteness, we consider the ascending leg of the loop, where $B_z>0$,
so that $0\leq \alpha \leq \pi/2$. The local curvature radius~$R(l)$ of the 
loop is related to~$\alpha(l)$ {\it via}
\beq
{1\over R} = {d\alpha\over{dl}} \, .
\label{eq-curvature-radius}
\eeq
Here we treat $\bar{B}(z)$ as a known function and our goal is to determine 
the geometrical shape of the loop described by~$\alpha(z)$.

The magnetic tension force on the loop segment acts in 
the direction of its curvature radius~$\hat{R}$ and is 
equal to
\beq
\delta f_{\rm tension} = {{B^2}\over{4\pi R}}\, a(l)\, \delta l \, ,
\eeq

The projection of the external-pressure force on the loop segment 
onto~$\hat{R}$ can be written as
%
%
\beq
\delta f_{\nabla P} = 
-\, {dP\over{dz}}\, \sin\alpha\ a(l)\, \delta l \, .
\eeq

Then, the force-balance condition can be written simply as
\beq
{B^2\over{4\pi}}\, {1\over R} = -\, \sin\alpha \, {dP\over{dz}} \, ,
\eeq
or, making use of equation~(\ref{eq-pressure-balance}), 
\beq
{{\bar{B}(z)}\over{R}} = -\sin\alpha \, {{d\bar{B}}\over{dz}} \, .
\label{eq-loopshape-1}
\eeq

Combining this equation with the geometrical relation~(\ref
{eq-curvature-radius}), we immediately obtain:
\beq
{d\alpha\over{dl}}={1\over R} = -\,{{d\log\bar{B}}\over{dz}}\, \sin\alpha\, .
\eeq
[Note that, as one can immediately see, a magnetic loop can be in equilibrium 
with a uniform [$\bar{B}(z)={\rm const}$] external pressure only if it is 
straight, $\alpha={\rm const}$.]

Using~(\ref{eq-def-alpha}), we get
\beq
\cot\alpha \, {d\alpha\over{dz}} = -\,{{d\log\bar{B}}\over{dz}} \qquad
\Rightarrow \qquad \sin\alpha(z) = {C\over{\bar{B}(z)}} \, .
\label{eq-alpha-Bbar}
\eeq

It is interesting to note that the same result can be obtained 
in a simple and elegant way by using a variational principle,
namely, by by minimizing the loop's magnetic energy~$E_{\rm magn}=
(\Delta\Psi/8\pi)\,\int B(l)\,dl$ (see \S~\ref{subsec-energy}) 
viewed as a functional of~$z(x)$. Using the relationship 
$dl = dx\, \sin^{-1}\alpha$, we have
\begin{eqnarray}
\delta E_{\rm magn}[z(x)] &=& 
{\Delta\Psi\over{8\pi}}\,\delta \int B[z(x)]\sqrt{1+z'^2(x)}\,dx \nonumber \\
&=& 
{\Delta\Psi\over{8\pi}}\,\int\biggl[B'(z)-B(z)\,{{z''}\over{1+z'^2}}\biggr]\, 
{{\delta z(x) dx}\over{\sqrt{1+z'^2}}} \, .
\end{eqnarray}
From the condition $\delta E_{\rm magn} = 0$, we thus immediately get
\beq
B = C \sqrt{1+z'^2(x)} = C {dl\over{dx}} = {C\over{\sin\alpha}} \, ,
\eeq
which, taking into account that $B=\bar{B}(z)$, is the same as 
the above result~(\ref{eq-alpha-Bbar}).

The result~(\ref{eq-alpha-Bbar}) means that the horizontal ($x$) component 
of the magnetic field is constant along the loop:
\beq
B_x(l) = B \sin\alpha = \bar{B}(z) \sin\alpha = C = {\rm const} \, .
\label{eq-B_x=B_top}
\eeq
The integration constant $C$ is just equal to the magnetic field~$B_{\rm top}$ 
at the top of the loop (where $\alpha=\pi/2$). Thus, the shape of the loop 
is given by the equation
\beq
\sin\alpha(z) = {B_{\rm top}\over{\bar{B}(z)}} \, .
\label{eq-loopshape-2}
\eeq

We can now work out an explicit expression for the field line shape~$z(x)$
in terms of the function~$b(z)\equiv\bar{B}(z)/\bar{B}(z=0)$. Let us 
denote the magnetic field strength at the base by $B_0\equiv \bar{B}(z=0)$ 
and the angle between the loop and the vertical direction at the base by 
$\alpha_0(L)=\alpha(z=0;L)$. We then have
\beq
b_{\rm top}(L) = {{\bar{B}_{\rm top}(L)}\over{B_0}} = \sin\alpha_0(L) \, .
\label{eq-b_top=sin-alpha_0}
\eeq

Then, the shape of the ascending leg of a loop of length $L$ is given by
\beq
x(z) = -\, b_{\rm top}\, \int\limits_z^{z_{\rm top}} \,
{{dz'}\over{\sqrt{b^2(z') - b^2_{\rm top}}}} =
-\, {L\over 2} + b_{\rm top}\, \int\limits_0^z \,
{{dz'}\over{\sqrt{b^2(z') - b^2_{\rm top}}}} \, .
\label{eq-loopshape-3}
\eeq

The height of the loop, $z_{\rm top}$, is determined implicitly by 
the condition $x(z_{\rm top})=0$:
\beq
{L\over 2} = b_{\rm top}\ \int\limits_0^{z_{\rm top}} \
{{dz'}\over{\sqrt{b^2(z') - b^2_{\rm top}}}}  \, .
\label{eq-z_top-1}
\eeq

Analogous expressions have been obtained by \cite{Parker75} 
and \cite{Browning_Priest84}.

Here are a few analytical examples of the use of this relationship.

1) Example I: 
\beq
b(z) = {1\over{1+\zeta}} \, ,
\label{eq-b(z)-example1}
\eeq
where $\zeta = z/z_0$; $z_0$ represents the characteristic magnetic 
scale height of the corona. 
Then we get $b_{\rm top}(L)= 2z_0\, (4 z_0^2+L^2)^{-1/2}$,
and $z_{\rm top}(L)=(z_0^2+L^2/4)^{1/2}-z_0$. 

2) Example II: exponential atmosphere,
\beq
b(z) = e^{-z/H}
\label{eq-b(z)-example2}
\eeq
Performing the integration~(\ref{eq-z_top-1}), we obtain 
$x(z)= -\, H\, \arctan\, [e^{2(z_{\rm top}-z)/H} - 1]^{1/2}$,
and correspondingly, 
\beq
L(Z_{\rm top}) = 2 H\, \arccos\,[{b(Z_{\rm top})}] =  
2 H\, \biggl[{\pi\over 2} - \alpha_0(Z_{\rm top})\biggr] \, .
\eeq

Notice that for tall loops with $Z_{\rm top}\gg H$ and $b_{\rm top}\ll 1$, 
the dependence $L(Z_{\rm top})$ saturates: $L(Z_{\rm top}) \rightarrow 
L_{\rm max} = \pi H$. This example illustrates an important point: if 
the external pressure drops off sufficiently steeply, then there is a 
maximum projected length, $L_{\rm max}$, that a loop in equilibrium can 
have. This implies that if one tries to insert a slender loop with a 
footpoint separation $L>L_{\rm max}$, then such a loop will not be able 
to attain an equilibrium and will instead grow in height without bound, 
{\it i.e.}, will tend to open up.

This fact points to an important feedback: for a given external pressure 
profile, large loops extend to larger heights, but this has an effect 
of increasing the contribution of these loops to the pressure at these 
large heights (see below), and hence may make the pressure profile less 
steep.


\subsection{Magnetic Energy Density~$\bar{B}^2(z)/8\pi$ 
as a Self-Consistent Mean Field}
\label{subsec-Bbar}

Our next step is to determine $\bar{B}$ as a function of~$z$.
However, the best way to do this is first to express $\bar{B}$
directly in terms of the length of the smallest loop $L$ that 
reaches the given height~$z$, {\it i.e.}, to find $\bar{B}(L)$. 
This can be done directly in terms of the orientation-integrated
distribution function $\bar{F}(L)\equiv \int F(L,\theta) d\theta$, 
since, at any given height~$z$, contributions to the magnetic 
pressure~$\bar{B}^2/8\pi$ come only from those loops that extend 
to this height or higher. Since the dependence~$Z_{\rm top}(L)$ 
is presumed to be monotonic, then $\bar{B}$ at a given height~$z$ 
will be proportional, roughly speaking, to the integral over all 
loops with lengths $L>L(z)$, where $L(z)$ is the function inverse 
to~$Z_{\rm top}(L)$. Thus, naively, we anticipate a result that 
looks something like this:
\beq
\bar{B}(L) \sim \Delta\Psi \int\limits_L^\infty \bar{F}(L)\, dL \, .
\eeq

In Appendix A we perform a rigorous analysis and derive an exact 
(within our model) result:
\beq
db = -\, {{\pi\,\Delta\Psi}\over{B_0}}\ \bar{F}(L)\, dL \quad\Rightarrow\quad 
b(L)={{\pi\,\Delta\Psi}\over{B_0}}\ \int\limits_L^\infty \bar{F}(L')\, dL' \,, 
\eeq
that is, 
\beq
B_{\rm top}(L) = \pi\,\Delta\Psi\ \int\limits_L^\infty \bar{F}(L')\, dL' \,.  
\label{eq-Btop(L)}
\eeq

The condition $b(z=0)=1$, in conjunction with $Z_{\rm top}(L=0)=0$, 
gives us the normalization condition for the function~$\bar{F}(L)$:
\beq
\int\limits_0^\infty \bar{F}(L') dL'  = {B_0\over{\pi\,\Delta\Psi}}  \, .
\label{eq-normalization}
\eeq
If all the loops were perpendicular to the disk surface, 
the normalization coefficient would be~1/2 (since each loop
has two footpoints). The extra factor~$(2/\pi)$ in the above 
expression reflects the fact that small low-lying loops are 
not perpendicular to the disk surface, and hence occupy larger 
horizontal projected area on this surface.

For a given loop distribution function~$\bar{F}(L)$, one can 
thus compute, in principle, the function~$b(L)$ that we shall 
need in the next subsection.


\subsection{Self-Consistent Loop Height $Z_{\rm top}(L)$}
\label{subsec-Ztop}

Once $L(b_{\rm top})$ is thus determined, we can substitute 
it into equation~(\ref{eq-z_top-1}) and thus reduce the whole 
problem to the following integral equation for the function~$U(b)
\equiv dz/db$:
\beq
{{L(b)}\over{2b}} = 
-\, \int\limits_b^1 {{U(b')\, db'}\over{\sqrt{b'^2 - b^2}}} \, .
\label{eq-Volterra-I}
\eeq
Once this equation is solved, we can integrate $U(b)$ to find~$z(b)$, 
thus completing the solution.

Mathematically, equation (\ref{eq-Volterra-I}) is a linear Volterra 
integral equation of the first kind for the function~$U(b)$ in terms 
of a known function~$L(b)$. It can be solved exactly. In particular, 
by a simple transformation of variables:
$t \equiv 1-b'^2$,
$s \equiv 1-b^2$, 
$G(s) \equiv L(b)/2b$, 
and $V(t) \equiv -\, U(b')/2b'$,
it can be transformed into the Abel equation:
\beq
\int\limits_0^s {{V(t)\,dt}\over\sqrt{s-t}} = G(s) \, .
\label{eq-Abel}
\eeq
whose solution is
\beq
V(s) = 
{1\over\pi}\, {d\over{ds}}\, \int\limits_0^s {{G(t)\,dt}\over\sqrt{s-t}} =
{1\over\pi}\, \int\limits_0^s {{G'(t)\,dt}\over\sqrt{s-t}} \, ,
\label{eq-Abel-soln}
\eeq
where we have used $G(s=0)=L(z=0)/2=0$.

Actually, the most useful form of the solution is 
the first equality in equation~(\ref{eq-Abel-soln}).
By substituting the definitions of~$V(s)$, $G(s)$ and~$U(b)$
into this equation, multiplying by $(-2b)$ and integrating 
with respect to~$b$, we get the following elegant final 
expression for the function~$z(b)$:
\beq
z(b) =  {1\over\pi} \int\limits_b^1 {{L(b')\,db'}\over\sqrt{b'^2-b^2}} \, .
\label{eq-z-of-b}
\eeq
Using equation~(\ref{eq-Btop(L)}), we can rewrite this 
in terms of the functions~$\bar{F}(L)$ and~$b(L)$ as 
\beq
Z_{\rm top}(L) =  {{\Delta\Psi}\over{B_0}} 
\int\limits_0^L {{L'\,\bar{F}(L')\,dL'}\over\sqrt{b^2(L')-b^2(L)}} \, .
\label{eq-Ztop-of-L}
\eeq


\subsection{Loop Energy}
\label{subsec-energy}

One of the main goals of this section is to address the {\it energetics} 
of the magnetized corona. Relevant issues include the energy distribution 
of flares as well as the torque on the disk due to the coronal magnetic 
fields. 
In order to be able to address this, we must first determine the energy, 
$\mathcal{E}(\mathcal{A})$, associated with a loop of type~$\mathcal{A}$.

The loop energy is given by the work done by the footpoints against 
magnetic forces as the loop's footpoint separation ({\it i.e.}, the 
projected length of the loop) is increased from zero to its present 
value~$L$. Since we regard $P(z)$ as isotropic, the energy depends 
only on the length but not on the orientation of the loop: 
\beq
\mathcal{E}(\mathcal{A}) = \mathcal{E}(L) =
\int\limits_0^L\, f_{\rm fp}(L')\, dL' \, . 
\label{eq-energy-1}
\eeq
Here, $f_{\rm fp}(L)$ is the magnetic force on each of the two footpoints;
it is proportional to the horizontal magnetic field~$B_{\rm hor}(z=0)$ at 
the disk surface:
\beq
f_{\rm fp}(L)={{B_z B_{\rm hor}}\over{4\pi}}\biggl|_{z=0}\,a_{\rm hor}(z=0)=
{\Delta\Psi\over{4\pi}}\, B_{\rm hor}(z=0;L) \, ,
\label{eq-f_footpt}
\eeq
where $\Delta\Psi$ is the magnetic flux carried by the loop and 
$a_{\rm hor}(z=0)=\Delta\Psi/B_z(z=0)$ is the area of the loop's 
footspot on the disk surface.%
\footnote
{Note that here, instead of fixing $\Delta\Psi$ and~$B_z(z=0)$,
we fix  $\Delta\Psi$ and the total magnetic field~$B(z=0)$, which
includes the horizontal component.}
Thus, we have
\beq
\mathcal{E}(L) =
{\Delta\Psi\over{4\pi}}\ \int\limits_0^L\,B_{\rm hor}(z=0;L')\, dL' \, .
\label{eq-energy-2}
\eeq

Another quantity of interest is the magnetic energy~$E_{\rm magn}$
contained within the loop:
\beq
E_{\rm magn} = \int a(l)\, {{B^2(l)}\over{8\pi}}\, dl \, ,
\label{eq-E_magn-1}
\eeq
where the integral is taken along the loop from one footspot to the other. 
Using flux conservation, $\Delta\Psi(l) = a(l) B(l) = {\rm const}$, this 
energy can be written simply as
\beq
E_{\rm magn} = {\Delta\Psi\over{8\pi}}\, \int B(l)\, dl \, .
\label{eq-E_magn-2}
\eeq
This expression actually has a very simple physical meaning. 
The integral $\int B(l)\, dl$ is the circulation of the magnetic 
field along the loop; according to the Ampere's law, this is 
just the total surface current flowing around the loop in the 
perpendicular direction. Thus, the above expression for $E_{\rm magn}$
is just a manifestation of the well-known result that the magnetic energy 
of a current circuit is proportional to the product of the magnetic flux 
enclosed by the circuit and its total current.

We would like to remark that $\mathcal{E}(L)$ can be viewed as 
a magnetic enthalpy, $H_{\rm magn}$. It includes both the magnetic 
energy~$E_{\rm magn}$ stored within the loop and the work~$W$ done 
by the loop on the surrounding gas with a fixed (but not necessarily 
uniform) pressure profile:
\beq
d\mathcal{E} = dH_{\rm magn} = dE_{\rm magn} + dW \, .
\eeq
This is analogous to calculating the amount of heat~$Q$ required 
to inflate a hot-air balloon at constant atmospheric pressure~$P_0$.
Indeed, when the balloon air is heated, the energy is expended both 
to increase the internal energy~$U$ of the hot air inside the balloon 
and to perform work against atmospheric pressure (neglecting heat losses 
from the balloon through its skin). Thus, the amount of heat that needs 
to be supplied is equal to the change in balloon's enthalpy~$H$:
\beq
dQ = dH = dU + P_0 dV = {1\over{\gamma-1}}\, d(P_0V) + P_0 dV =
{\gamma\over{\gamma-1}}\, P_0 dV \, ,
\label{eq-balloon}
\eeq
where $\gamma$ is the adiabatic index of air. In our case of a magnetic loop
confined by external pressure, the adiabatic index is~$\gamma=2$. Therefore,
we expect that the total energy that needs to be supplied by the footpoint 
motions to inflate the loop is twice the internal magnetic field energy within 
the loop:
\beq
\mathcal{E} = E_{\rm magn} + W = 2\, E_{\rm magn} \, .
\eeq
In Appendix B we demonstrate that this is indeed so.

Note that, in addition to the work $W$ done against the external 
gas pressure as the loop expands and increases its cross-section,
there is also the work done against the magnetic tension force as 
the length of the loop is increased and the work done on the loop 
by the external pressure-gradient force. However, as long as the loop 
expands quasi-statically, always maintaining its equilibrium shape, 
the last two forces precisely balance each other (see \S~\ref{subsec-shape}),
and so their corresponding works cancel out.

Using the formalism developed in the previous subsections,
we can now easily calculate the energy associated with a 
given loop. Substituting equation~(\ref{eq-B_x=B_top}) into 
equation~(\ref{eq-energy-2}) and using expression~(\ref{eq-Btop(L)}) 
for $B_{\rm top}(L')$, we get
\beq
\mathcal{E}(L)={\Delta\Psi\over{4\pi}} \int\limits_0^L\,B_{\rm top}(L')\,dL'= 
{{\Delta\Psi^2}\over{4}}\ 
\int\limits_0^L \int\limits_{L'}^\infty \bar{F}(L'')\, dL'' \, dL' \, .
\label{eq-energy-3}
\eeq

The total magnetic energy in the corona is then
\begin{eqnarray}
E_{\rm tot} &=& 
{1\over 2}\ \int\limits_0^\infty\ \bar{F}(L)\,\mathcal{E}(L) \, dL =
{{\Delta\Psi^2}\over{8}} \ 
\int\limits_0^\infty dL \int\limits_0^L dL' \int\limits_{L'}^\infty dL'' 
\ \bar{F}(L)\, \bar{F}(L'')  \nonumber \\
&=& {{\Delta\Psi^2}\over{4}} \ 
\int\limits_0^\infty dL \int\limits_0^L dL'\ L'\, \bar{F}(L)\, \bar{F}(L') \, .
\label{eq-E_tot}
\end{eqnarray}

It is instructive to consider a case in which $\bar{F}(L)$ has a power law 
tail, $\bar{F}\sim L^{-\alpha}$, truncated at some large $L_{\rm max}\gg H$. 
Then, as can be seen from equation~(\ref{eq-energy-3}), the energy of 
the largest loop, $\mathcal{E}(L_{\rm max})$, is almost independent of
$L_{\rm max}$ for $\alpha>2$, but starts to grow as~$L_{\rm max}^{2-\alpha}$ 
for $\alpha<2$. 
In a similar manner, from equation~(\ref{eq-E_tot}) it follows that
the total coronal energy becomes dominated by the large-$L$ tail if 
$\bar{F}(L)$ drops off sufficiently slowly, {\it i.e.}, $\alpha<3/2$. 
In this case the total energy scales with~$L_{\rm max}$ as $E_{\rm tot}
\propto L_{\rm max}^{3-2\alpha}$ and may become much larger than the 
potential magnetic field (whose characteristic magnetic scale height
is of order~$H$). Physically, we expect $L_{\rm max}$ to be at most 
about the local disk radius, $r$, so that $E_{\rm tot}$ is bounded.

Finally, let us consider the angular momentum transfer by the coronal
magnetic field. The torque due to a single loop is given by
\beq
\Delta G(L,\theta) = 
-\, {{\Delta\Psi}\over{4\pi}}\ |B_{\rm hor}|\ {{\Delta y\,\Delta r}\over L}= 
-\, {{\Delta\Psi B_0}\over{8\pi}}\ b_{\rm top}(L)\ L\, \sin{2\theta} \, ,
\label{eq-loop-torque}
\eeq 
and hence the total torque per unit disk area is
\beq
G = -\, {{\Delta\Psi B_0}\over{8\pi}}\ \int\int dL\, d\theta\ 
\bar{F}(L,\theta)\, b_{\rm top}(L)\ L\, \sin{2\theta} \, .
\label{eq-total-torque}
\eeq 

Let is again consider the truncated power-law example, $\bar{F}\sim 
L^{-\alpha(\theta)}$. For a fixed degree of anisotropy, {\it e.g.}, 
a fixed the characteristic angular scale~$|\sin{\theta_{\rm min}}|$ 
at which the function $\alpha(\theta)$ has a minimum~$\alpha_{\rm min}$, 
the torque becomes dominated by large loops if $\alpha_{\rm min}<3/2$,
similar to~$E_{\rm tot}$ [where we used equation~(\ref{eq-Btop(L)}].
It then scales as~$L_{\rm max}^{3-2\alpha}$ and thus may become by 
a factor $(L_{\rm max}/H)^{3-2\alpha}\gg 1$ larger than the usual  
torque exerted directly by MRI turbulence within the disk, assuming
that the magnetic field at the base of the corona is comparable to 
that in the disk. In reality, however, a decrease in $\alpha_{\rm min}$ 
may come hand-in-hand with an increase in the degree of anisotropy of 
the distribution function~$\bar{F}(L,\theta)$ (see \S~\ref{subsec-results}),
manifested as a simultaneous decrease in $|\sin{\theta_{\rm min}}|$.
If this is the case, the torque amplification will be not as strong.



\section{The Loop Kinetic Equation}
\label{sec-kinetic-eqn}

In this section we discuss how to calculate the loop distribution function.
In particular, we construct the loop kinetic equation that governs the 
evolution of this function.


\subsection{Physical Assumptions of our Model}
\label{subsec-assumptions}

In order to build a quantitative model of the magnetic field 
in the corona, we need to make some specific assumptions about 
the most important physical processes that govern the life of 
individual coronal loops, including their interactions with each 
other. These assumptions are the main building blocks of our model;
we shall discuss them in this section.

1) The Alfv\'en velocity in the corona is much faster than both the disk's 
rotational velocity at the given radius and the thermal velocity of the 
coronal gas; therefore, the corona is considered to be in a slowly-evolving 
force-free magnetostatic equilibrium at all times and almost everywhere 
(except for rapid rearrangements due to reconnection events, see below).

2) The disk is geometrically {\it thin}, with the gas scale-height 
much smaller than the distance from the central object, $H\ll r$. 
This gives us an important small parameter that can be used in the 
analysis. For example, this assumption gives us an ``inertial range'' 
of spatial scales much smaller than~$r$ but much larger than~$H$. 
This enables us to perform an analysis that is local in~$r$. 
In particular, this means that we can neglect geometrical effects 
resulting from cylindrical geometry when considering the flux-loop 
expansion process. Note, however, that the validity of the thin-disk 
assumption is questionable close to a black hole accreting near its
Eddington limit.

3) The disk is {\it differentially-rotating} ({\it e.g.}, keplerian).
As a result, coronal loops with radially-separated footpoints
are subject to continuous stretching in the toroidal direction.
This generates toroidal magnetic field whose pressure inflates
the loops and ultimately leads to the creation of a vertically-extended 
corona (see below).

4) At any given time, the shape and the overall height of each loop 
are determined by the magnetostatic equilibrium of the loop as if it 
were confined by a stratified atmosphere with certain external isotropic 
pressure~$P(z)$. In turn, this pressure represents the effective magnetic 
pressure of all the neighboring loops, and we shall denote it as 
$P=\bar{B}^2(z)/8\pi$. This equilibrium shape is maintained at all 
times, since it adjusts on the Alfv\'en time scale, which is assumed to 
be much faster than the disk rotation.

5) The disk is {\it turbulent} due to the usual internal MRI 
(as opposed to coronal that may act simultaneously). 
The characteristic spatial scale of the turbulence is~$H$,
and the characteristic time scale is~$\Omega_K$.
The important effects of the disk turbulence on the corona are:

5a) {\it Flux Emergence} plays a very important role 
in the solar corona and, by analogy, is also believed 
to be important in the case of the disk \cite{Galeev_Rosner_Vaiana79}. 
We generally expect the emerging magnetic loops to be relatively small 
in size (of order the disk thickness~$H$) and to have typical magnetic 
fields of order~$\alpha_{\rm ss}^{1/2} B_{\rm eq}$, where the dimensionless
parameter $\alpha_{\rm ss}\sim 0.01-0.1$ is the \cite{Shakura_Sunyaev73}
viscosity coefficient and $B_{\rm eq}$ is the field strength that 
corresponds to equipartition with the gas pressure inside the disk. 
In addition, numerical studies of MRI turbulence 
show that the toroidal field in the disk tends to be larger than 
the radial field by a factor of 5-10. Thus, flux emergence is expected 
to be anisotropic, with newly emergent loops elongated in the toroidal 
direction by a factor of a few.

5b) Again, similar to the Sun, the disk turbulence leads to a 
two-dimensional random walk of the coronal loops' footpoints
on the disk surface. We expect this random walk to be characterized
by spatial and temporal scales of the order of~$H$ and~$\Omega^{-1}$, 
respectively.  
However, similar to the process of flux emergence discussed above, 
the random walk, in general, may be anisotropic, with characteristic 
steps in the azimuthal direction being somewhat larger than in the 
radial direction.

\begin{figure}[t]
\plotone{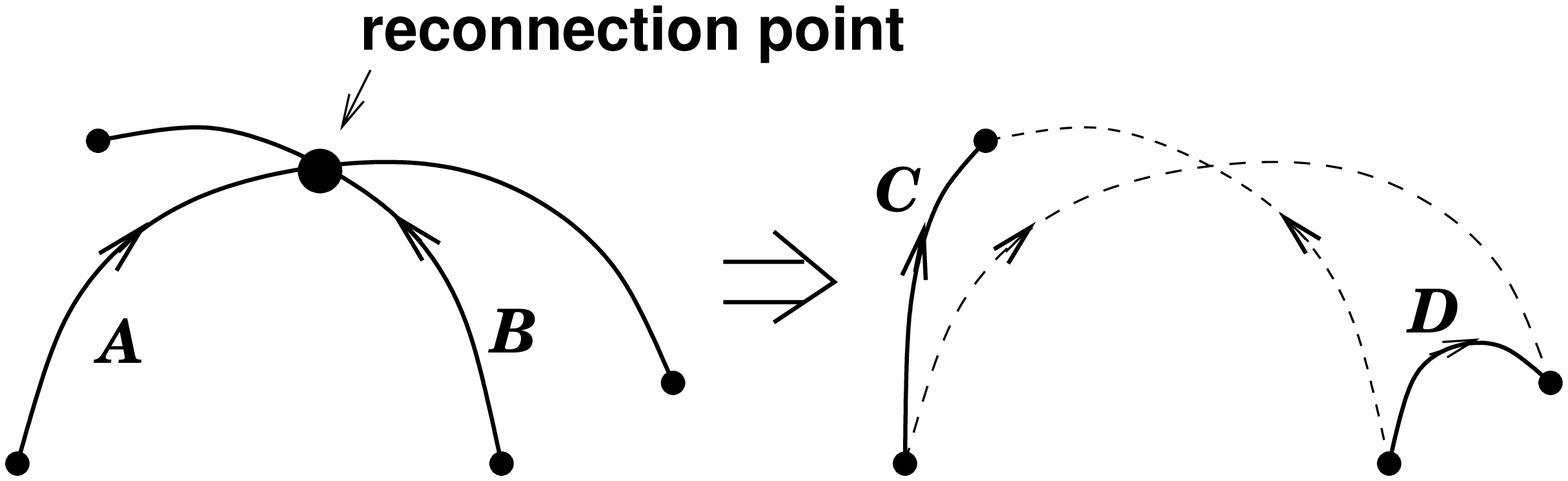}
\figcaption{Reconnection between two loops as a binary process.
\label{fig-recn}}
\end{figure}

6) {\it Reconnection:}  
In our model, two loops may {\it reconnect} with each other, forming 
two new loops (see Fig.~\ref{fig-recn}). Thus, reconnection represents 
a binary interaction between individual magnetic structures, analogous 
to binary collisions between particles in a gas. 

We shall assume that, once triggered, a reconnection event (a flare)
happens very quickly, essentially instantaneously on the orbital time 
scale. This assumption can be justified by noting that the corona is 
assumed to be a very low-density, and hence collisionless, environment. 
Therefore, reconnection there proceeds in the Petschek-like fast 
collisionless regime, enabled by anomalous resistivity or by the 
two-fluid (Hall-MHD) effects. 
The characteristic reconnection time-scale is then only by a factor 
of 10-100 slower than the Alfv\'en crossing time~$\tau_A$.
Thus, since we assume that $V_A \gg V_K$, it is reasonable to expect 
that the typical duration of coronal reconnection events may still be 
fairly short compared with the orbital time-scale~$\Omega^{-1}$. 
Then, to the extent that~$\Omega^{-1}$ is the main dynamical time-scale 
in our problem, characterizing differential rotation, flux emergence, 
and turbulent random walk, we can, for the purposes of our study, 
regard reconnection between loops, once triggered, as being essentially 
instantaneous. 
Thus, we arrive at a picture in which magnetic loops evolve slowly
({\it i.e.}, on the  orbital time scale), but from time to time they 
suddenly and instantaneously reconnect. This picture is similar 
to the observed behavior of solar coronal loops, where the characteristic 
reconnection (or flare) time is typically much shorter than the typical 
loop lifetimes. Thus, from the standpoint of viewing the corona as an 
ensemble of many loops, reconnection events can be regarded as relatively 
infrequent binary collisions between loops, analogous to the binary 
collisions between particles in Boltzmann's gas. 
An important corollary from this is that the footpoints of the loops 
do not have time to move significantly during the reconnection event. 
As we shall see in~\S~\ref{subsec-LKE}, this will give us the rules 
that determine the footpoint separations of newly-formed loops. 
We shall also assume that these newly-formed loops quickly assume 
their equilibrium shapes (see above).

\subsection{The Loop Kinetic Equation}
\label{subsec-LKE}

Based on the above assumptions, we shall now work out an 
evolution equation for the loop distribution function~$F$. 
We shall call this the {\it Loop Kinetic Equation} (LKE). 
This equation should have the following terms, reflecting 
the relevant physical processes:

1) Flux emergence/submergence acts as a source/sink of new coronal loops. 
It can be modeled by a source term, $S(\mathcal{A})$, that describes the 
rate at which the loops emerge into the corona, their characteristic sizes, 
magnetic field strengths, etc., (or, in a more elaborate model, by specifying 
the distributions of these quantities). Specifically, one can add loops at 
some characteristic ``injection scale'', somewhat larger than the disk 
thickness~$H$, and remove very small loops, say, of size~$H$ or less,
as it was done in the model by \cite{Hughes_Paczuski_etal03}. 
In addition, in the ADC case, we expect flux emergence to be anisotropic, 
with the emerging loops being by a factor of a few longer in the toroidal 
direction than in the radial direction, as indicated by numerical 
simulations \cite[\emph{e.g.}][]{Hirose_Krolik_Stone06}.

An alternative way to take flux emergence into account is {\it via} 
the boundary conditions for $F$ at small scales (of order~$H$). This 
is the view adopted in our model. This choice is justified by arguing 
that the population of smallest loops comprising the ``magnetic carpet'' 
is predominantly determined by a detailed balance that is quickly 
established with the magnetic fields in the disk itself. This process 
turns over (operates) very quickly and hence the distribution of the 
very small loops is basically independent of what happens in the 
larger-scale corona.

2) Random footpoint motions due to the disk turbulence. Since we expect
the characteristic steps of this random walk to be relatively small 
(of order~$H$, see \S~\ref{subsec-assumptions}) compared with the sizes 
of most loops under consideration, we can employ a Fokker--Planck-like 
approach to this process. This results in a diffusion operator 
with the diffusion coefficient of the order of the \cite{Shakura_Sunyaev73} 
$\alpha$-viscosity coefficient: $D\simeq \alpha_{\rm ss} c_s H \sim
\alpha_{\rm ss} \Omega H^2$. 
In general, however, this diffusion may be anisotropic, with~$D$ 
being a tensor ({\it e.g.}, a diagonal tensor with $D_{yy}> D_{rr}$).
We expect the effect of the random walk to be relatively unimportant 
for large loops, $L\gg H$.

3) Keplerian differential rotation leads to a secular evolution of~$\Delta y$, 
\beq
{d\Delta y\over{dt}} = -\, {3\over 2}\, \Omega\, \Delta r \, .
\label{eq-shear}
\eeq
Here, $\Omega = \Omega_K(r)$ can be regarded as constant because 
we consider spatial scales that are small compared with the disk 
radius, $\Delta r \ll r$.
In the Loop Kinetic Equation this process is described by an advection 
term, $(3/2)\, \Omega\, \Delta r \, (\partial F/\partial\Delta y)$.

4) Coronal MRI backreaction term~$\dot{F}_{\rm backreaction}$: 
in a geostrophic approximation, this is obtained by balancing 
the magnetic force on the footpoints (per unit area) with the 
Coriolis force (also per unit area) due to the rotation of the 
loop it induces: 
\beq
2\, {{B_z {\bf B}_{\rm hor}}\over{4\pi}} = 
2\,  \Sigma\,  [{\bf \Omega} \times \delta {\bf v}_{\rm backreaction}] \, ,
\eeq
in which ${\bf v}_{\rm backreaction}$ represents the departure 
from the keplerian rotation velocity.

5) Interaction of two loops by reconnecting with each other,
yielding two new loops. This process is described by a binary 
collision term~$\dot{F}_{\rm rec}$.
This is the most non-trivial term and we devote the entire 
next subsection (\S~\ref{subsec-recn-term}) to a detailed 
discussion of it.

At the end, we arrive at the following general form of the kinetic equation 
for the loop distribution function:
\begin{eqnarray}
{\partial F\over{\partial t}}(\Delta r,\Delta y,t) &=&
S(\Delta r,\Delta y) +
\biggl( D_{rr}\,{\partial^2\over{\partial \Delta r^2}} + 
D_{yy}\,{\partial^2\over{\partial \Delta y^2}} \biggr) F \nonumber \\
&+& {3\over 2}\,\Omega \Delta r\,{{\partial F}\over{\partial\Delta y}} +
\dot{F}_{\rm backreaction} + \dot{F}_{\rm rec}
\label{eq-LKE}
\end{eqnarray}

The simplest meaningful case of this equation is when one neglects 
the source, diffusion, and feedback terms and looks for a {\it steady 
state} that is produced by the balance between Keplerian shear and
reconnection:
\beq
{3\over 2}\, \Omega \Delta r\, 
{{\partial F}\over{\partial \Delta y}} = \dot{F}_{\rm rec} \, .
\label{eq-steady-state}
\eeq


Our approach will thus be analogous to, and can be regarded as an 
extension of, the previous work by \cite{Tout_Pringle96}, whose 
main goal was to study the formation of large magnetic structures 
{\it via} the reconnective ``inverse cascade'' in the corona. 
Following them, we also represent the coronal magnetic field by 
an ensemble of flux loops described by a distribution function. 
However, our model is more general and uses more realistic physics. 
We take into account a number of effects ignored in \cite{Tout_Pringle96},
such as inflation of the loops as they are stretched by the Keplerian
differential rotation. Also, in their model reconnection was taking place 
only at the disk surface and thus one of two newly reconnected loops was 
vanishingly small and was assumed to just disappear; as a result,
the reconnection process did not conserve the number of loops.
We, on the other hand assume that reconnection occurs higher in 
the corona, and hence two new loops form and the loop number is 
conserved (similar to the model by \cite{Hughes_Paczuski_etal03}).



\subsection{Reconnection Described as a Collision Integral}
\label{subsec-recn-term}

Two loops $\mathcal{A}$ and~$\mathcal{B}$ may interact by reconnecting 
with each other and forming two new loops~$\mathcal{C}$ and~$\mathcal{D}$ 
as a result (see Fig.~\ref{fig-recn}). Following \cite{Tout_Pringle96}, 
we shall describe this process by a nonlinear binary-collision integral, 
similar to the collision integral in the Boltzmann kinetic equation. 
In reality, of course, interaction between loops is more complicated 
and so such a description is oversimplified. Moreover, magnetic loops 
fill up the entire coronal space, and so they resemble more a non-ideal 
liquid rather than an almost ideal rarefied gas with infrequent binary 
encounters. Nevertheless, we believe that this binary-collision representation 
of reconnection can lead to some valuable physical insight into the 
complicated dynamics of the coronal magnetic field.

The Boltzmann collision integral can be split into two: the source term and 
the sink term. The sink term $\dot{F}_{\rm coll,-}(\mathcal{A})$ describes 
the rate of reduction in the number of loops of a given type~$\mathcal{A}$ 
due to reconnection between these loops and all other loops. 
The source term $\dot{F}_{\rm coll,+}(\mathcal{A})$ describes 
the rate of increase in the number of loops of type $\mathcal{A}$ 
when they are a product of reconnection of other-type loops. 
By ``type'' we here mean a set of loops with the same values 
of their primary parameters $(L,\theta)$ or $(\Delta r,\Delta y)$.
Thus, each of these terms is a quadratic integral operator, with 
a kernel that depends both on the types of the two loops and also 
on their relative position (see below). Thus, we can write the 
reconnection term schematically as
\beq
\dot{F}_{\rm rec}(\mathcal{A}) = 
\dot{F}_{\rm rec,-}(\mathcal{A}) + \dot{F}_{\rm rec,+}(\mathcal{A}) \, ,
\label{eq-Fdot_rec}
\eeq
where
\begin{eqnarray}
\dot{F}_{\rm rec,-}(\mathcal{A}) &=& - \int d\mathcal{B}\ 
Q_{\mathcal{A}\mathcal{B}} \, F(\mathcal{A}) \, F(\mathcal{B})\, , 
\label{eq-Fdot-}            \\
\dot{F}_{\rm rec,+}(\mathcal{A}) &=& 
{1\over 2}\, \int\int d\mathcal{C}\, d\mathcal{D}\ 
Q_{\mathcal{C}\mathcal{D}\rightarrow\mathcal{A}} \
F(\mathcal{C})\,F(\mathcal{D})\, .
\label{eq-Fdot+}   
\end{eqnarray}
Here, $d\mathcal{B}\equiv d\Delta r_B\, d\Delta y_B$, etc.,
The kernels $Q_{\mathcal{A}\mathcal{B}}$ in the sink term and 
$Q_{\mathcal{C}\mathcal{D}\rightarrow\mathcal{A}}$ in the source 
term are related {\it via}
\beq
Q_{\mathcal{A}\mathcal{B}} = {1\over 2}\, \int d\mathcal{C}\ 
Q_{\mathcal{A}\mathcal{B}\rightarrow\mathcal{C}} \, .
\eeq 

Using this relationship, the two terms can be combined as
\beq
\dot{F}_{\rm rec}(\mathcal{A}) = 
{1\over 2}\, \int \int d\mathcal{C}\, d\mathcal{D}\ F(\mathcal{D})\, 
\biggl[\, Q_{\mathcal{C}\mathcal{D}\rightarrow\mathcal{A}}\, F(\mathcal{C})- 
Q_{\mathcal{A}\mathcal{D}\rightarrow\mathcal{C}}\, F(\mathcal{A})\, \biggr]\, .
\label{eq-Fdot_rec-2}
\eeq

In order to go from this general expression to a specific 
operational procedure, we need to formulate the rules that 
govern the reconnection process. Indeed, the two new loops 
$\mathcal{C}$ and~$\mathcal{D}$ formed as products of reconnection 
between two loops~$\mathcal{A}$ and~$\mathcal{B}$ cannot be arbitrary 
and certain selection rules must be applied. More specifically, from 
the assumption that reconnection is instantaneous on the orbital time 
scale, it follows that the footpoints of the reconnecting loops do 
not move significantly during the reconnection event. Only the way 
they are connected to each other changes. Therefore, the primary 
parameters ({\it i.e.}, footpoint separations) of loops~$\mathcal{C}$ 
and~$\mathcal{D}$ are uniquely determined by the footpoint positions 
of the two incoming loops~$\mathcal{A}$ and~$\mathcal{B}$:
\begin{eqnarray}
\Delta r_{C} &\equiv & r_{C-} - r_{C+} = r_{A-} - r_{B+} \, , 
\label{eq-rec-rule-C}                 \\
\Delta r_{D} &\equiv & r_{D-} - r_{D+} = r_{B-} - r_{A+} \, ,
\label{eq-rec-rule-D}
\end{eqnarray} 
and similarly for $\Delta y_{C}$, $\Delta y_{D}$.
Here, $A_+$, $A_-$ are the positive and negative polarity footpoints 
of the loop~$\mathcal{A}$, {\it etc}. Mathematically, these rules play 
a role similar to the momentum and energy conservation conditions 
for particle collisions in kinetic theory of gases; they enter as 
$\delta$-functions in the interaction integral in our loop kinetic 
equation. Thus, one can easily see that the parameters of the new 
loops depend not only on the parameters of the old loops, but also 
on the positions of~$\mathcal{A}$ and~$\mathcal{B}$ relative to each 
other (see below). 

The kernel $Q_{\mathcal{A}\mathcal{B}}$ in equation~(\ref{eq-Fdot-}) is 
the probability rate ({\it i.e.}, probability per unit time) that two loops 
of types~$\mathcal{A}$ and~$\mathcal{B}$ will come together and reconnect. 
Thus, $Q_{\mathcal{A}\mathcal{B}}$ describes the rate of reconnection events 
(the number of such events per unit time). This should not be confused with 
the concept of ``reconnection rate'', a widely-used term in reconnection 
research with a completely different meaning.

Now let us discuss on which physical parameters~$Q_{\mathcal{A}\mathcal{B}}$ 
should generally depend. Whereas Tout \& Pringle (1996) just took 
$Q_{\mathcal{A}\mathcal{B}}={\rm const}$, we want to develop a more 
realistic and more sophisticated model, taking into account several 
important factors.
First, notice that~$Q_{\mathcal{A}\mathcal{B}}$ has dimensions of 
$[{\rm cm}^2/{\rm sec}]$. Based on dimensional arguments, it should 
then be proportional to the characteristic rate at which the coronal 
magnetic field is reconfigured. In addition, it should also reflect 
the fact that larger loops have larger ``interaction cross-section'' 
(see below), and thus should be roughly proportional to the squares 
of loop sizes.

Let us first address the characteristic reconfiguration time-scale.
The only fundamental dynamical time-scale in the corona, {\it i.e.}, 
the time-scale on which the corona, seen as an ensemble of elementary 
coronal structures, rearranges itself, is the orbital time, $\Omega^{-1}$
(or the inverse of the shear rate, $3/2\Omega$, which is not independent). 
This means that, if we represent the evolution by a sequence of discrete 
steps, each step representing a noticeable change in the relative position 
and/or orientation of the coronal elements, then the most appropriate choice 
for duration of these  steps is of order~$\Omega^{-1}$. Therefore, in general, 
$Q$ should scale with~$\Omega$. Next, if we follow a given magnetic element, 
at each new step there will be a certain probability $\kappa<1$ that the 
resulting new magnetic configuration around this element becomes favorable 
for reconnection of this element with another. We shall treat~$\kappa$ as 
a constant number, independent of the loops primary parameters. Thus, the 
overall rate at which the loops are disrupted through reconnection with 
other elements should be proportional to~$\kappa\Omega$:
\beq
Q_{\mathcal{A}\mathcal{B}} = \kappa\, \Omega\, \sigma_{AB} \, ,
\eeq
where we introduced the ``reconnection cross-section''~$\sigma_{AB}$.

The cross-section~$\sigma_{AB}$ should in some way scale with the loop sizes.
It involves contributions from all possible relative positionings of 
the two interacting loops for which the two loops ``effectively intersect''.
We shall describe this relative positioning by two impact parameters, 
$b_\parallel$ and~$b_\perp$, defined as the offsets between the centers 
of the two loops in the direction parallel and perpendicular to 
loop~$\mathcal{A}$, respectively. We shall assume that once the impact 
parameters are in a range such that the two loops ``effectively intersect'', 
the probability that these loops will reconnect is constant, independent 
of their positions or their parameters. Furthermore, for simplicity we 
shall assume~$b_\parallel$ and~$b_\perp$ to be uniformly-distributed 
independent random variables. Thus, $\sigma_{AB}$ is just equal to 
the ``interaction area'' in the $(b_\parallel,b_\perp)$ space that 
corresponds to an ``effective intersection'' of the given two loops:
\beq
\sigma_{AB} = \int d\sigma_{AB} =  \int\int  db_\parallel \, db_\perp \, .
\eeq

But what do we mean by ``effectively intersecting''?
If the two loops are approximated by their central lines
(one-dimensional objects), then the set of values $(b_\parallel,b_\perp)$
for which they intersect is also one-dimensional (a line segment), 
and thus has measure zero in the two-dimensional $(b_\parallel,b_\perp)$ 
space. In other words, the probability that two randomly-drawn lines 
intersect is zero in the three-dimensional space. Therefore, 
to get a meaningful result, we need to take into account finite 
thicknesses of the loops. In particular, we shall say that two 
loops ``effectively intersect'' when the closest distance between 
their central lines is less than a certain fraction of the combined 
loop thicknesses at the intersection height. 
Operationally, for a given value of~$b_\parallel$, say, we can introduce 
the $y$-cross-section $\sigma_{AB,\perp}(b_\parallel)$ as the spread in 
the values of~$b_\perp$ which result in an effective intersection of two 
given loops~$\mathcal{A}$ and~$\mathcal{B}$. We can then write 
\beq
\sigma_{AB} = \int \sigma_{AB,\perp}(b_\parallel)\, db_\parallel \, . 
\label{eq-sigma_perp}
\eeq
where the integral is taken over the range of impact parameters~$b_\parallel$
for which an intersection between loops~$\mathcal{A}$ and~$\mathcal{B}$ 
is at all possible.

Correspondingly, the sink-term part of the reconnection integral
can be written as
\beq
\dot{F}_{\rm rec,-}(\mathcal{A}) = 
-\, \kappa\Omega\ \int\int d\mathcal{B}\ db_\parallel\,
\sigma_{\mathcal{A}\mathcal{B},\perp}(b_\parallel)\, 
F(\mathcal{A})\, F(\mathcal{B})\, .
\label{eq-Fdot-2}
\eeq


Now let us consider the source term~(\ref{eq-Fdot+}). 
Employing the arguments given earlier in this subsection, we 
introduce the cross-section for two given loops~$\mathcal{C}$ 
and~$\mathcal{D}$ to reconnect giving a loop $\mathcal{A}$ as 
a result: $Q_{\mathcal{C}\mathcal{D}\rightarrow\mathcal{A}}=
\kappa\,\Omega\,\sigma_{CD\rightarrow A}$. We can then write
\beq
\dot{F}_{\rm coll,+}(\mathcal{A}) = 
{{\kappa\Omega}\over 2}\ \int\int d\mathcal{C}\, d\mathcal{D}\, 
\sigma_{\mathcal{C}\mathcal{D}\rightarrow\mathcal{A}}\ 
F(\mathcal{C})\, F(\mathcal{D}) \, .
\label{eq-Fdot+2}
\eeq

The $\mathcal{D}$ integral in this equation is taken over all the loops 
that can yield a loop of type~$\mathcal{A}$ as a result of reconnection 
with a loop of a given type~$\mathcal{C}$. Note that, whereas one did not 
need to know reconnection product loops to compute the sink term~(\ref
{eq-Fdot-2}), in order to calculate the source term, this knowledge is 
in fact necessary. It is contained in our ``reconnection rules'', such 
as those given by equations~(\ref{eq-rec-rule-C})--(\ref{eq-rec-rule-D}).
For definiteness, let us consider the case when the resulting 
loop~$\mathcal{A}$ starts from the ``+'' footpoint of loop~$\mathcal{C}$
and ends at the ``-'' footpoint of loop~$\mathcal{D}$. 
Then, for given~$\mathcal{C}$ and~$\mathcal{A}$, there is a well-defined 
range of impact parameters~$b_\parallel$ for which one can find one 
(or sometimes two) loop~$\mathcal{D}=\mathcal{D}_{\mathcal{A}-
\mathcal{C}}(b_\parallel)$ that intersects loop~$\mathcal{C}$ 
and ends at the end footpoint of loop~$\mathcal{A}$. Thus, in 
principle, for a given~$b_\parallel$ one can formulate the rules 
that relate the primary parameters~$\Delta r_D$ and~$\Delta y_D$ 
to those of loops~$\mathcal{C}$ and~$\mathcal{A}$. We shall denote these 
relationships by $\Delta r_D^{A-C}(b_r)$ and $\Delta y_D^{A-C}(b_\parallel)$. 
In general, there may be one or two such solutions.

Recalling now that flux tubes have a finite thickness, 
we have, by analogy with equation~(\ref{eq-sigma_perp}),
\beq
\sigma_{CD\rightarrow A} = 
2\, \int db_\parallel\, \sigma_{CD,\perp}(b_\parallel)\
\delta[\mathcal{D}-\mathcal{D}_{\mathcal{A}-\mathcal{C}}(b_\parallel)] \, ,
\eeq
where 
$\delta[\mathcal{D}-\mathcal{D}_{\mathcal{A}-\mathcal{C}}(b_\parallel)] \equiv
\delta[\Delta r_D - \Delta r_D^{A-C}(b_\parallel)]\, \times
\delta[\Delta y_D - \Delta y_D^{A-C}(b_\parallel)]$.
The factor~2 accounts for the fact that in the preceding paragraph 
we considered only one half of all possible configurations, requiring 
the starting footpoint of loop~$\mathcal{A}$ to be the starting footpoint 
of loop~$\mathcal{C}$. For each such configuration there will also 
be an identical contribution from interchanging loops~$\mathcal{C}$ 
and~$\mathcal{D}$.

Substituting this cross-section into our expression~(\ref{eq-Fdot+2}) 
for~$\dot{F}_{\rm coll,+}(\mathcal{A})$, and using the $\delta$-function 
to integrate over~$d\mathcal{D}=d\Delta r_D d\Delta y_D$, we get
\beq
\dot{F}_{\rm coll,+}(\mathcal{A}) = 
\kappa\, \Omega \int \int d\mathcal{C}\, db_\parallel\ 
\sigma_{CD_{A-C},\perp}(b_\parallel)\, 
F(\mathcal{C})\, F[\mathcal{D}_{\mathcal{A}-\mathcal{C}}(b_\parallel)]\, .
\label{eq-Fdot+3}
\eeq

Combining equations~(\ref{eq-Fdot-2}) and~(\ref{eq-Fdot+3}), we can write:
\begin{eqnarray}
\dot{F}_{\rm rec}(\mathcal{A}) &=& 
\kappa\, \Omega\ \int\int d\mathcal{C}\ db_\parallel\, F(\mathcal{C})\,
\nonumber\\
&\times& \biggl[F[\mathcal{D}_{\mathcal{A}-\mathcal{C}}(b_\parallel)] 
\sigma_{CD_{A-C},\perp}(b_\parallel) - F(\mathcal{A})
\sigma_{AC,\perp}(b_\parallel)\biggr] .
\label{eq-Fdot_rec-3}
\end{eqnarray}


\subsubsection{Effect of Finite Loop Thickness}

For a given value of~$b_\parallel$, one first finds the coordinates 
$[r(b_\parallel),y(b_\parallel),z(b_\parallel)]$ of the point where 
the central lines of the two loops would intersect. After that, one 
calculates the $\perp$-extent of each of the two loops at this point, 
which hence gives one~$\sigma_{AB,\perp}(b_\parallel)$.

When doing this, one should take into account the following important effect. 
If the loops have no internal twist, as we assume here, the longitudinal 
magnetic field is approximately constant 
across a loop and is roughly equal to the characteristic magnetic 
field $\bar{B}(z)$ at a given height~$z$. 
Then, by flux conservation, the cross-section of the loop (normal 
to its central line) varies along its length, and, in particular, 
may increase greatly at large heights if $\bar{B}(z)$ drops off rapidly.
More precisely, approximating a loop's cross-section at a given height~$z$ 
as a circle of some radius~$d(z)$, we can estimate this radius as
$d(z)\sim [\Delta\Psi/\pi\bar{B}(z)]^{1/2}= d_0 [\bar{B}(z)/B_0]^{-1/2}$.
As long as the loop is slender, $d\ll L$, this should be a good approximation.
Thus, if the intersection point lies high above the disk, so that 
$\bar{B}(z)\ll B_0$, the characteristic loop thickness is much larger 
than near the disk surface. Consequently, the reconnection cross-section 
is increased, which has important implications for the ``inverse 
cascade'' of magnetic loops in the corona. It is also interesting to 
note that this process is controlled by the self-consistent field~$\bar{B}(z)$.

[Note that, in the case of twisted loops this effect is not as profound
and a more accurate approximation is probably given by 
$d(z)\simeq d_0={\rm const}$ (see~\S~\ref{subsec-twisted}).]

Thus, typically we expect $\sigma_{AB,\perp}$ to scale as 
\beq
\sigma_{AB,\perp}(b_\parallel) \sim 2\, d[z(b_\parallel)] 
\sim 2\, \sqrt{{\Delta\Psi}\over{\pi\bar{B}[z(b_\parallel)]}} 
\sim 2\, d_0 \sqrt{B_0\over{\bar{B}[z(b_\parallel)]}} \, .
\label{eq-sigma_ABperp-2}
\eeq

For example, if $\mathcal{B}$ is the smaller of the two loops, 
we expect the typical interaction height to be of the order of 
this loop's height~$Z_B$. 
Correspondingly, we expect $\sigma_{AB,\perp} \sim d_0 
b^{-1/2}(Z_B)$, where $b(z)\equiv \bar{B}(z)/B_0$.
This estimate for $\sigma_{AB,\perp}(b_\parallel)$ will be roughly 
valid for almost the entire allowed range of~$b_\parallel$, 
which is of order $4\,\Delta r_B$. 

To sum up, larger loops have larger cross-sections for reconnection, 
for two complementary reasons. First, the cross-section is enhanced 
because larger loops have larger range of impact parameters 
(in the radial direction, say) for which intersection of 
their central lines is possible. Second, larger loops 
extend to, and may interact at, larger heights, where 
the mean magnetic field is weaker and hence where the 
loops become fatter. They thus have a greater chance 
of overlapping with each other. As a result, the reconnection 
cross-section scales with the size~$L_B$ of the {\it smaller} 
of the two loops as 
\beq
\sigma_{AB} \sim L_B\, d_0\, b^{-1/2}(Z_B)\, ,
\eeq
[for $L_B\gg d(Z_B)$]. We thus see that the function~$\bar{B}(z)$
affects the evolution of the loop distribution function. Since, 
according to~\S~\ref{subsec-Bbar}, $\bar{B}(z)$ is itself determined 
by the distribution function, this means that determining these two
functions together in a self-consistent way requires an iterative 
procedure.

What is important here, is that larger loops have a tendency to 
reconnect with each other quickly, probably leading to a rapid 
``inverse cascade'' to even larger loops. In addition, the fact 
that reconnection with large loops cannot be neglected suggests
that a Fokker-Planck-like approximation to the reconnection term
will not work. This is because reconnection events that lead to 
large changes in loop parameters are important and so our collision
integral cannot be described by a differential diffusion-like 
operator.


\section{Numerical Solution}
\label{sec-numerical}

\subsection{Numerical Setup}
\label{subsec-setup}

We solve the LKE numerically and obtain a steady state solution.
For simplicity we leave out the turbulent diffusion term, the coronal 
backreaction term, and the source term (flux emergence is then treated 
{\it via} the boundary condition at small scale, see \S~\ref{subsec-bc}).
Thus, we include only the two processes which we believe 
are the most important: Keplerian shear and reconnection
between loops. Correspondingly, we aim here at investigating 
the effect of the relative importance of these two processes 
on the steady-state loop distribution function.

In the numerical implementation we work in the $(L,\theta)$ parameter
space where $L=(\Delta r^2+\Delta y^2)^{1/2}$ is the distance between 
the footpoints and $\theta$ is the angle the vector ($\Delta r,\Delta y$)
makes with the toroidal direction, measured clockwise:
$\tan\theta = \Delta r/\Delta y$. 
We use a grid that is uniform in $\theta$ (between 0 and $2\pi$) 
but logarithmic in~$L$. The ${L}$~grid spans from $L_{\rm min}=1$ 
to some $L_{\rm max}\gg 1$ (usually we take $L_{\rm max}$ to be~10 
or~20) in length units such that $H\simeq 1$.

The advection term resulting from Keplerian shear is very easy to implement, 
we just use one-sided derivatives.

The reconnection collision integral is obviously more complicated and we 
devote the rest of this subsection to our numerical implementation of it. 
At each time step we go over all possible pairs of loops~$\mathcal{A}$ 
and~$\mathcal{B}$. Furthermore, for each given pair we go over all 
possible reconnecting configurations, distinguished by the impact 
parameter~$b_\parallel$, defined as the displacements between the 
centers of the two loops in the direction along loop~$\mathcal{A}$ 
(see \S~\ref{subsec-recn-term}). Thus, at each timestep, we are 
performing a five-dimensional (5D) integration, which makes increasing 
resolution extremely numerically costly.
We found, however, that numerical convergence is very good and 
relatively modest resolution suffices. To study numerical convergence, 
we performed calculations with $N_L=20$ and~40 points in~$\log{L}$, 
$N_\theta=20$, 40, 60, and~80 points in~$\theta$, and 30, 50, or~100
points in~$b_\parallel$. We found the resulting~$F(L,\theta)$ to 
be essentially unchanged, although small values of $\kappa$ required 
a higher $\theta$-resolution for convergence (see below).

For simplicity, in our treatment of reconnection, we assume the loops 
to be {\it semi-circular} in shape, instead of using equilibrium shapes 
discussed in~\S~\ref{subsec-shape}. This simplification has two benefits. 
First, since the shapes are described by relatively simple analytical 
expressions, we can derive explicit analytical relationships expressing 
the parameters of the two product loops in terms of the parameters of 
the interacting loops and the impact parameter (we call these 
relationships the  ``reconnection rules''). Having such expressions
in an explicit form greatly simplifies the numerical procedures.

The second advantage of the semicircular approximation stems from 
the observation that, because all the loops have the same shape, the 
reconnection rules essentially depend only on the ratio of loop sizes,
whereas their dependence on the absolute loop size is a trivial rescaling.
Likewise, the reconnection process is, by itself, isotropic, {\it i.e.}, 
does not depend on the absolute orientation of the loops, only on the 
angle between them.
This enables us to to reduce the analysis of reconnection between 
two given loops  $\mathcal{A}$ and~$\mathcal{B}$  to considering
a template that corresponds to the given angle between the loops 
and the given length ratio~$L_B/L_A$. That is, we can analyze 
reconnection between any two loops in a rotated and rescaled 
system of coordinates, $(x',y')$, in which one of the loops 
(loop $\mathcal{A}$ for definiteness) is in the positive $x'$-direction, 
{\it i.e.}, has $\theta_A'=\pi/2$ and has unit length, $L_A'=1$.
In practice, we first (even before we start the evolution of~LKE), 
create a lookup table describing the reconnection rules for interaction 
of this loop~$\mathcal{A}'$ with all other loops~$\mathcal{B}'$.
The lookup table is~3D, two of the coordinates being~$\theta_B'$ 
and~$L_B'$, and the third coordinate being the rescaled impact 
parameter~$b_\parallel'$; the latter lies within the range from 
$-(1+\Delta x_B')/2$ to $(1+\Delta x_B')/2$, where $\Delta x_B'=
L_B'\sin\theta_B'$. For each given~$\theta_B'$, $L_B'$ and~$b_\parallel'$,
it is just a matter of simple algebra to figure out the perpendicular 
displacement $b_\perp'$ corresponding to the intersection between the 
semi-circular loops. 
Simultaneously one finds the 3D position of the intersection point between 
loops~$\mathcal{A'}$ and~$\mathcal{B'}$ for given~$b_\parallel'$ (including 
the height of the reconnection point). One can then readily deduce the 
parameters of the two product loops~$\mathcal{C'}$ and~$\mathcal{D'}$.
It is easy to see that the problem reduces to quadratic equations and hence 
for each~$\theta_B'$, $L_B'$, and~$b_\parallel'$ there may be zero, one, or 
two solutions. 

During the time evolution we use this table as follows. For each given 
pair of loops~$\mathcal{A}$ and~$\mathcal{B}$ we find their corresponding 
template pair by rotating them by $\theta_A-\pi/2$ and rescaling the loop 
sizes by~$L_A$. We also rescale the impact parameter by~$L_A$, {\it i.e.}, 
$b_\parallel'=b_\parallel/L_A$.
We then use the template table to find the two template product loops 
and we transform them back by multiplying by~$L_A$ and rotating by 
$\pi/2-\theta_A$ to find the actual product loops~$\mathcal{C}$ 
and~$\mathcal{D}$. Having a grid uniform in~$\theta$ and~$\log{L}$ 
makes this procedure especially convenient and straightforward.

Once the product loops are found, one also needs to figure out
the reconnection cross-sections corresponding to given $\mathcal{A}$, 
$\mathcal{B}$, and~$b_\parallel'$. In accordance with the above
discussion, $d\sigma_{AB}(b_\parallel')$ is proportional to 
$d b_\parallel=L_A\,(1+\Delta x_B')/N_b)$ and to the loop combined 
deprojected thickness~$db_\perp'$ at the intersection point. Apart 
from simple geometrical projection factors, $db_\perp'$ is proportional 
to $d(z) \sim d_0 [B_0/\bar{B}(z)]^{1/2}$, where~$z$ is the height of 
the reconnection point above the disk. This factor enhances the 
reconnection probability for large loops intersecting at large 
heights. Since $\bar{B}(z)/B_0$ itself depends on~$F(L,\theta)$, 
this procedure requires iteration.
Once the reconnection cross-section $\sigma_{AB}(b_\parallel')$ is 
found, one can proceed to evolve the number of incoming~($\mathcal{A}$ 
and~$\mathcal{B}$) and product ($\mathcal{C}$ and~$\mathcal{D}$) loops. 
Namely, $F(\mathcal{A})d\mathcal{A}$ and~$F(\mathcal{B})d\mathcal{B}$ 
are reduced by 
\beq
F(\mathcal{A})\, F(\mathcal{B})\, d\mathcal{A}\, d\mathcal{B}\ 
L_A\, db_\parallel'\ \kappa\, \Omega\,\sigma_{AB}(b_\parallel')\, \Delta t \,,
\eeq 
and $F(\mathcal{C})\,d\mathcal{C}$ and $F(\mathcal{D})\,d\mathcal{D}$ 
are increased by the same amount at each timestep~$\Delta t$.

Finally, we varied the initial conditions and found that our resulting 
steady state solutions are insensitive to them.


\subsection{Boundary Conditions}
\label{subsec-bc}

For various physical reasons, we expect our model to break down 
at both small scales and large scales. We thus need to discuss how 
to prescribe the boundary conditions for the~LKE at both of these
scales.

At small scales, $L\sim H$, the model is expected to become invalid
because the magnetic field is no longer force-free, as plasma pressure
starts to become dynamically important at small heights. In addition,
the characteristic thickness of magnetic structures near the disk
surface is expected to be of order~$H$ and hence small loops with $L\sim H$ 
are not going to remain slender as our model assumes. Thus we need to 
understand what is a plausible way to describe these small loops. 
At small scales the dynamics of magnetic loops is strongly affected 
by rapid flux-exchange ({\it e.g.} flux-emergence) processes with the 
turbulent disk. One can therefore argue that the overall distribution 
of the smallest loops is largely determined by the detailed balance 
equilibrium that is rapidly established between smallest loops 
and the turbulent disk.  Characterizing this intense interaction 
with the disk turbulence and calculating the small-scale distribution 
function is completely beyond our present model. It would require thorough 
understanding of 3D MHD turbulence in stratified disks with simultaneous 
actions of the MRI and Parker instability \citep{Tout_Pringle92}; 
most likely, this task will be accomplished using sophisticated numerical 
simulations. However, what matters for our model here is that we expect 
that the number of small loops be essentially insensitive to what happens 
to larger coronal loops. 
This means that we can mimic the disk-corona interaction by setting-up 
a Dirichlet-type boundary condition at small scales, {\it i.e.}, by 
prescribing the distribution function at some small-cutoff scale 
$L_{\rm min}\sim H$:
\beq
F(L=L_{\rm min},\theta) = F_1(\theta) \, .
\eeq
(In our model we set $L_{\rm min}=1$.)
Notice that $F_1(\theta)$ is in general expected not to be isotropic
because emerging loops will be preferentially azimuthally elongated 
($|\sin\theta|\ll 1$). In the present simulations, however, we take
it to be isotropic, $F_1(\theta)={\rm const}$.

Our model also breaks down at large scales. In particular, when a loop's 
size becomes comparable to the disk radius~$r$, the local cartesian geometry 
adopted here is no longer applicable, and toroidal effects become important. 
This leads to a much faster expansion of stretched loops, resulting in a 
complete opening of the field, as discussed in~\S~\ref{subsec-recn-role}.
In the future, we plan to implement a physically realistic way of 
treating the opening of large loops by incorporating open field lines 
into our model (see~\S~\ref{subsec-recn-collisionless}).

In the present model, however, we just introduce some large cut-off 
scale, $L_{\rm max}$, and we set boundary conditions at this scale. 
We experimented with two types of boundary conditions:
in one case, we set $F(L_{\rm max})=0$, {\it i.e.}, we simply remove all 
loops that reach the cut-off scale. In the other case, we just limit the 
growth of loops beyond $L_{\rm max}$; {\it i.e.}, as soon as a loop's 
length exceeds~$L_{\rm max}$, we reset it back to $L_{\rm max}$. 
This leads to a gradual pile-up of loops near~$L_{\rm max}$, but does 
not affect the loop distribution on smaller scales; in particular, it 
does not change the power law tail for sizes just slightly smaller 
than~$L_{\rm max}$.


\subsection{Results}
\label{subsec-results} 

\subsubsection{Distribution Function}

\begin{figure}[h]
\plotone{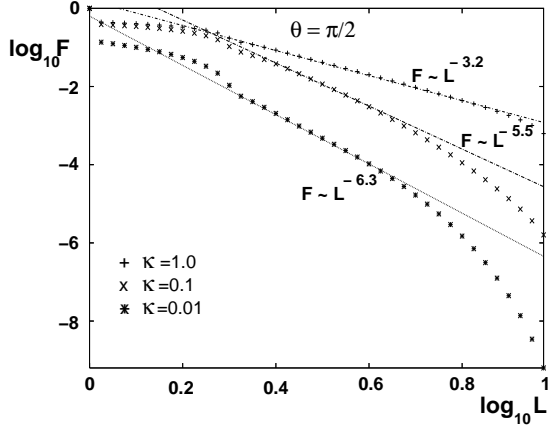}
\figcaption{Distribution function for purely radial loops ($\theta=\pi/2$),
for dimensionless reconnection parameter $\kappa=1.0$ (pluses), 0.3 (crosses), 
and~0.1 (asterisks).
\label{fig-radial-F}}
\end{figure}

\begin{figure}[h]
\plotone{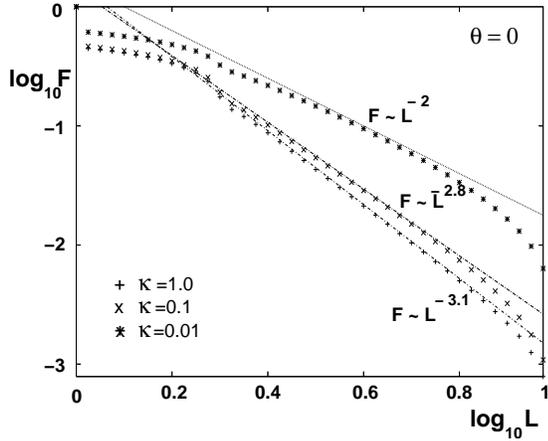}
\figcaption{Distribution function for purely toroidal loops ($\theta=0$),
for $\kappa=1.0$ (pluses), 0.1 (crosses), and~0.01 (asterisks).
\label{fig-azimutal-F}}
\end{figure}

\begin{figure}[h]
\plotone{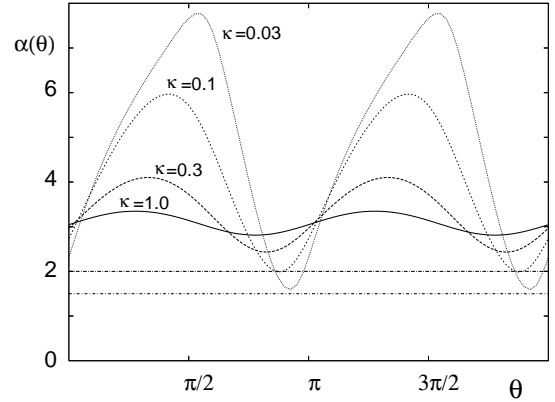}
\figcaption{Power-law exponent $\alpha_\kappa(\theta)$ for 
$\kappa=1.0$, 0.3, 0.1, 0.03.
\label{fig-alpha-theta}}
\end{figure}

We performed a series of calculations with several different values of~$\kappa$
($\kappa=\infty$, 1, 0.3, 0.1, 0.03, 0.02, 0.01, 0.005, and~0.003). Some 
of the resulting steady state loop distribution functions for our fiducial 
resolution $N_L=40$, $N_{\theta}=80$ and for the large-scale boundary 
condition $F(L_{\rm max}, \theta)=0$, $L_{\rm max}=10$, are presented 
in Figures~\ref{fig-radial-F}--\ref{fig-alpha-theta}. 
In particular, Figures~\ref{fig-radial-F} and~\ref{fig-azimutal-F} 
show (in log-log coordinates) $F$ as a function of~$L$ for purely 
radial loops ($\theta=\pi/2$) and purely toroidal loops ($\theta=0$), 
respectively. 
In the case $\kappa=\infty$ (keplerian shear turned off), the distribution 
function is isotropic. This is of course expected, since loop-loop 
reconnection --- the only process determining the distribution function 
in this case --- by itself is independent of the absolute orientation of 
the reconnecting loops. As the frequency of reconnection events relative 
to shear, quantified by~$\kappa$, is decreased, a given loop (especially
if it has a large~$|\Delta r|$) experiences, on average, larger stretching 
in the toroidal direction by the shear before it undergoes reconnection 
with another loop. As a result, the loops become predominantly azimuthal
and the loop distribution function becomes more and more anisotropic: 
$F(\theta=\pi/2,L)$ steepens, whereas $F(\theta=0,L)$ becomes shallower 
with decreasing~$\kappa$.

Because  our problem lacks a preferred length-scale between~$L_{\rm min}$
and~$L_{\rm max}$, we find that, along each ray $\theta={\rm const}$, 
$F(L)$ is well described by a power law with the orientation-angle-dependent 
exponent:
\beq
F(L,\theta) \sim L^{-\alpha_\kappa(\theta)} \, .
\eeq
Figure~\ref{fig-alpha-theta} presents the function~$\alpha_\kappa(\theta)$ 
for $\kappa=1$, 0.3, 0.1, and~0.03.

\begin{figure}[h]
\plotone{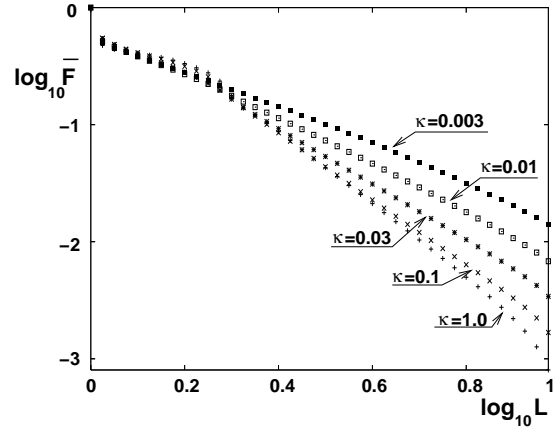}
\figcaption{Orientation-averaged loop distribution function~$\bar{F}(L)$ 
for $\kappa=1.0$, 0.3, 0.1, 0.03, 0.01, and~0.005.
\label{fig-averaged-F-q}}
\end{figure}

Finally, in order to estimate the total magnetic energy in the corona
and the total magnetic torque (see~\S~\ref{subsec-energy}), one needs 
to know the $\theta$-integrated distribution function, 
\beq
\bar{F}(L) \equiv \int\limits_0^{2\pi} F(L,\theta)\, d\theta \, .
\eeq
In Figure~\ref{fig-averaged-F-q} we plot $\bar{F}(L)$ for $\kappa=1.0$, 
0.3, 0.1, 0.03, 0.01, and~0.005 in log-log coordinates. In general, of 
course, one cannot expect a $\theta$-integral of exponents 
$L^{-\alpha(\theta)}$ to be itself a power law of~$L$.
However, we find that the integral is strongly dominated by the range
of values~$\theta$ corresponding to the ridge in~$F(L,\theta)$, and so 
a power law $\bar{F}(L)\sim L^{-\bar{\alpha}(\kappa)}$ is actually a 
reasonably good approximation, especially for relatively large values 
of~$\kappa$. Because the power-law fit is not ideal, there is some 
degree of uncertainty in determining the value of~$\bar{\alpha}$; 
we estimate the characteristic error to be~$\lesssim 0.1$.

\begin{figure}[h]
\plotone{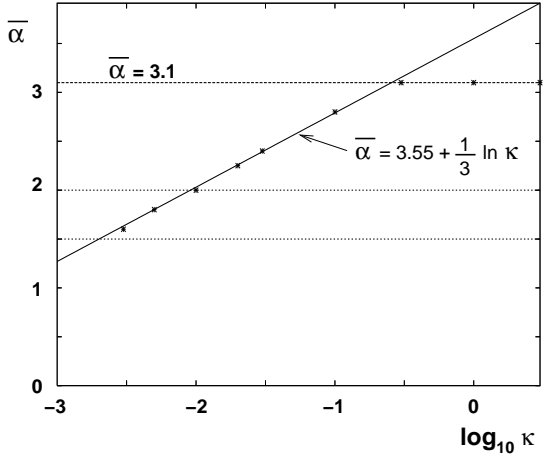}
\figcaption{Power-law exponent $\bar{\alpha}$ for the orientation-averaged 
loop distribution function~$\bar{F}(L)\sim L^{-\bar{\alpha}}$, as a function 
of~$\kappa$. 
The dotted horizontal lines at $\bar{\alpha}=2$ and $\bar{\alpha}=3/2$ 
correspond, respectively, to the critical values at which the magnetic 
energy of the largest loop, $\mathcal{E}(L_{\rm max})$, and the total 
magnetic energy in the corona, $E_{\rm tot}$, start to become dominated 
by the large-scale cut-off~$L_{\rm max}$ (see~\S~\ref{subsec-energy}).
\label{fig-alpha_bar-q-log}}
\end{figure}

The dependence of the power-law exponent~$\bar{\alpha}$ on the reconnection 
parameter~$\kappa$ is plotted in Figure~\ref{fig-alpha_bar-q-log}.
We find that when reconnection is strong compared with keplerian shear 
($\kappa \gtrsim 0.2$), the power-law exponent stays close to~3.1, 
independent of~$\kappa$. This is because, as long as the keplerian 
shear can be neglected, reconnection is the only term on the right-hand
side of the~LKE. In this case, ~$\kappa$ cannot affect the steady state 
solution, it can only regulate how fast this solution is achieved.
As~$\kappa$ is decreased, however, the keplerian shear term becomes
important and $\bar{\alpha}(\kappa)$ starts to decrease. 
We find that its overall behavior can be approximated by 
$\bar{\alpha}(\kappa)\simeq 3.1 = {\rm const}$ for~$\kappa\gtrsim 0.2$ 
and $\bar{\alpha}(\kappa)\simeq 3.55+(1/3)\, \ln{\kappa}$ for~$\kappa
\lesssim 0.2$. Correspondingly, we expect $\bar{\alpha}(\kappa)$ to 
crosses the critical values~2 and~3/2 (see~\S~\ref{subsec-energy}) 
at $\kappa_2\simeq 0.01$ and~$\kappa_{3/2} \simeq 0.002$, respectively.
As a reminder, $\bar{\alpha}<3/2$ means that the total magnetic energy 
of the corona is dominated by the largest loops.

Finally, although most of our runs were done with $L_{\rm max}=10$, 
we also conducted some runs with~$L_{\rm max}=20$. We found the 
distribution function to be essentially the same; the power-law 
just extended further in the $L_{\rm max}=20$ case but the slope 
and the normalization were unchanged.


\subsubsection{Energetics and Torque}

\begin{figure}[h]
\plotone{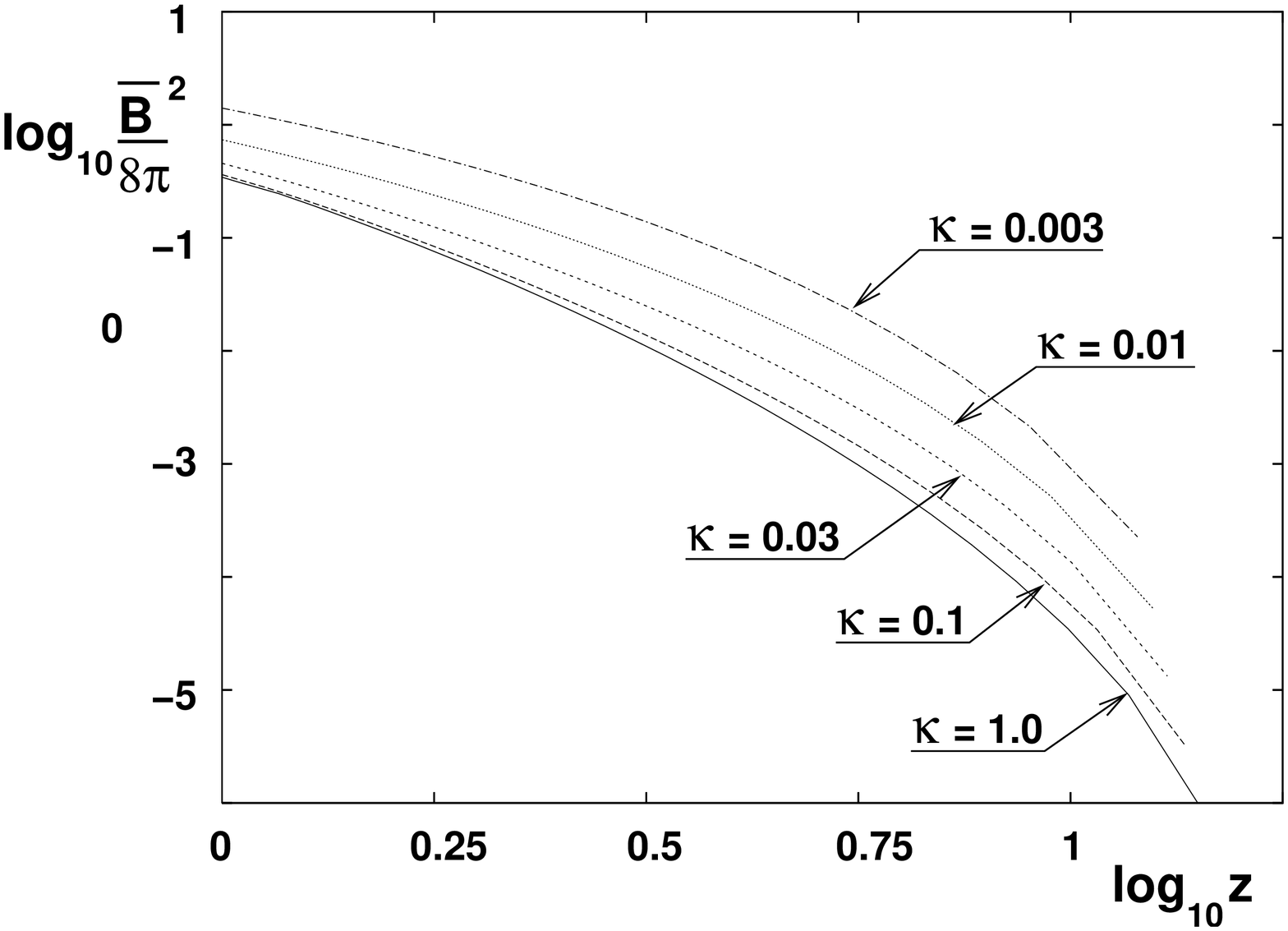}
\figcaption{Magnetic energy density as a function of height, 
$\bar{B}^2(z)/8\pi$, for several values of the dimensionless 
reconnection parameter~$\kappa$.
\label{fig-Bbar-squared-z}}
\end{figure}

\begin{figure}[h]
\plotone{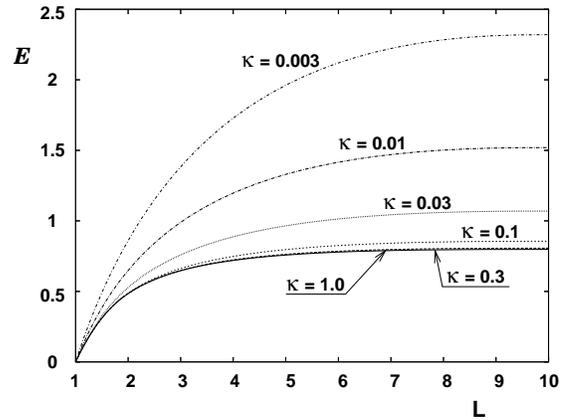}
\figcaption{Energy associated with a loop of length~$L$, $\mathcal{E}(L)$, 
for several values of the dimensionless reconnection parameter~$\kappa$.
\label{fig-Energy-L}}
\end{figure}

Using the computed loop distribution functions, we calculate 
the magnetic energy density as a function of height above the disk, 
$\bar{B}^2(z)/8\pi$, and the energy of a loop as a function of its
length, $\mathcal{E}(L)$, for several different values of~$\kappa$ 
(see Figs.~\ref{fig-Bbar-squared-z} and~\ref{fig-Energy-L}).
We see that these functions remain essentially independent of~$\kappa$ 
as long as it is large enough, $\kappa \gtrsim 0.1$), but start changing
for smaller values of~$\kappa$. In particular, we find that for a given
value of~$\kappa$, $\mathcal{E}(L)$ saturates to a finite value at large~$L$.
This is because, although a large loop occupies a relatively large volume 
at large~$z$, the magnetic energy density very high above the disk is very 
low, and so the contribution of the large-$z$ part of a loop to its total
energy is relatively small.
However, we find that the asymptotic value of~$\mathcal{E}(L)$ at large~$L$ 
begins to increase rapidly as~$\kappa$ is lowered below roughly~$\kappa_2
\simeq 0.01$, in agreement with the expectations of~\S~\ref{subsec-energy}.

\begin{figure}[h]
\plotone{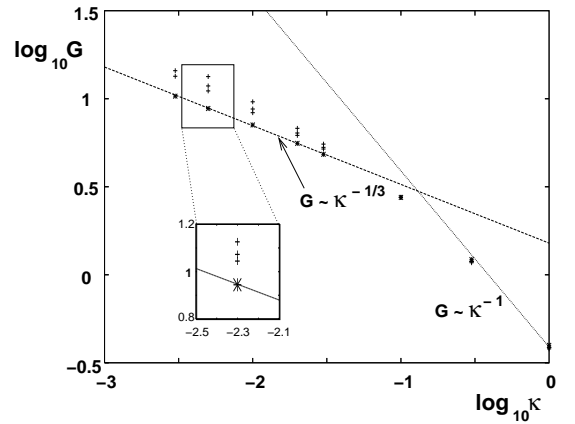}
\figcaption{Coronal magnetic torque $G$ (normalized to $\Delta\Psi^2/4\pi$)
as a function of~$\kappa$ in logarithmic coordinates. The small insert shows 
an expanded view of convergence with respect to the resolution in the~$\theta$
direction for $\kappa=0.005$ ($\log_{10}{\kappa}=-2.3$). The plus signs mark 
the values obtained with $N_\theta=40$, 60, and 80 (top to bottom) and the 
asterisks correspond to the extrapolation to~$N_\theta=\infty$.
\label{fig-torque-of-q-log}}
\end{figure}

\begin{figure}[h]
\plotone{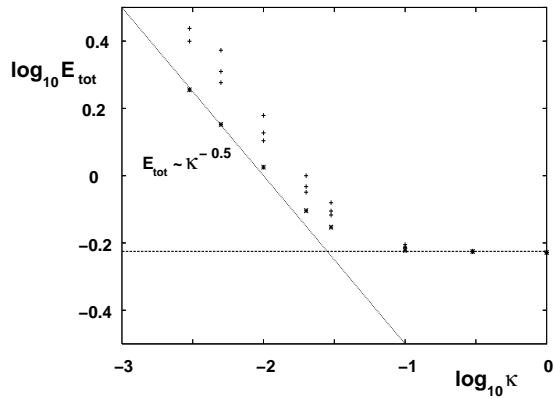}
\figcaption{Total coronal magnetic energy (normalized to~$\Delta\Psi^2/4$)
as a function of~$\kappa$ in logarithmic coordinates. As in Figure~\ref
{fig-torque-of-q-log}, the plus signs correspond to $N_\theta=40$, 60, 
and 80 (top to bottom) and the asterisks corresponds to the extrapolations 
to $N_\theta=\infty$, for each~$\kappa$.
\label{fig-Etot-of-q-log}}
\end{figure}

We also calculate the total torque~$G$, transmitted by the coronal 
magnetic field and the total magnetic energy in the corona, $E_{\rm tot}$. 
We show these quantities as functions of~$\kappa$ in Figures~\ref
{fig-torque-of-q-log} and~\ref{fig-Etot-of-q-log}. 
Since our code is only first-order accurate with respect to 
the number~$N_\theta$ of grid points in the $\theta$~direction, 
the finite $\theta$-resolution becomes an issue at very small 
values of~$\kappa$, where the distribution function is strongly 
anisotropic. Therefore, to get more accurate values for the torque 
and the magnetic energy corresponding to~$N_\theta=\infty$, we use 
linear extrapolation based on the calculations with $N_\theta=40$, 
60, and~80, separately for each value of~$\kappa$ (see Figs.~\ref
{fig-torque-of-q-log} and~\ref{fig-Etot-of-q-log}).

We find that, as $\kappa$ is decreased, the coronal angular 
momentum transfer~$G(\kappa)$ increases steadily as~$\kappa^{-1}$ 
for $\kappa\gtrsim 0.1$, as is expected due to the increased degree 
of anisotropy of the loop distribution function. 
As~$\kappa$ becomes $\lesssim 0.1$, further growth of~$G(\kappa)$ slows
down and becomes $G\sim \kappa^{-1/3}$. The fact that $G(\kappa)$ is 
a power law in this range is understandable in light of the discussion 
at the end of~\S~\ref{subsec-energy}, combined with the logarithmic 
dependence of~$\bar{\alpha}(\kappa)$ reported above.
We also find that the total magnetic energy in the corona, 
$E_{\rm tot}(\kappa)$, is nearly flat for large values of~$\kappa$, 
but starts to increase roughly as~$\kappa^{-1/2}$ as~$\kappa$ is 
lowered below $\kappa\lesssim 0.05$ (see Fig.~\ref{fig-Etot-of-q-log}).

Furthermore, in agreement with the discussion in~\S~\ref{subsec-energy},
we find that both the magnetic torque and the total magnetic energy are 
substantially larger in the $L_{\rm max}=20$ case than in the $L_{\rm max}=10$ 
case for small values to~$\kappa$, 
whereas they are essentially the same for both values of~$L_{\rm max}$ 
for larger~$\kappa$.


\section{Discussion}
\label{sec-discussion}

There are several extensions of our model that we plan to develop in the near
future. They include:  (1) characterizing the coronal field backreaction on 
the disk motions; (2) including open magnetic flux; (3) taking into account 
magnetic twist inside the loops; (4) incorporating a more realistic 
prescription for reconnection; (5) investigating the mass exchange 
between the disk and the corona and its effect on regulating coronal 
energy release; (6) studying the interaction between the coronal magnetic 
field and a super-imposed external large-scale field; 
(7) assessing the implications of our theory for observations, 
{\it e.g.}, in terms of time delay between hard and soft X-ray emission. 
We discuss some of these issues in more detail in this section. 



\subsection{Open Flux Tubes, Outflows, and Net Vertical Flux}
\label{subsec-open}

In this paper we have assumed that closed loops are the only magnetic
structures in the corona. In principle, however, one should also
consider a population of open flux tubes.  If considered to be
force-free along their entire (infinite) length, open loops exert 
no torques on the disk (in the non-relativistic limit) but contribute 
to the averaged magnetic pressure~$\bar{B}^2(z)$ everywhere and thus 
prevent it from becoming too small at large heights.  Open tubes are 
also essential for investigating the role of the corona in launching 
large-scale outflows; relaxing the assumption of force-free fields 
allows the inertia of the outflow to exert a torque on the disk 
along open field lines \citep{Blandford_Payne82} by analogy with 
the angular-momentum loss of the Sun to the solar wind \citep{Weber_Davis67}.

In our model, open loops could be introduced through the large-scale
cut-off for the closed loops. That is, whenever a closed loop grows to
exceeds a certain maximal footpoint separation~$L_{\rm max}$, it could
be replaced by a pair of open flux tubes of opposite polarities. 
The large-scale cut-off may be physically associated with the disk
radius,~$r$. It should determine the fraction of magnetic flux that 
is open at any given time. The rules that govern the interaction of 
open tubes with closed ones and with themselves, are straightforward: 
two open tubes of the same sign do not interact; two open tubes of
opposite sign can annihilate by reconnection, forming one single
closed loop. An open tube can also reconnect with a closed loop,
forming again an open tube of the same sign and a new closed loop.

In addition to such pairs of positive and negative open flux tubes,
there may also be a net vertical magnetic flux imposed on the disk 
by the central object or the ambient inter-stellar medium. 
The radial transport of such externally-imposed field across a turbulent 
accretion disk is an important problem with significant consequences for 
understanding the different spectral states of accreting black holes
\citep{Spruit_Uzdensky05}, for production and collimation of disk-driven 
winds and jets, and for star--disk magnetic interaction \citep[\emph{e.g.}]
[]{Uzdensky_Koenigl_etal02a,Uzdensky_Koenigl_etal02b}, which is believed 
to regulate the spin evolution of accreting neutron stars in X-ray pulsars
\citep{Ghosh_Lamb78}, as well as young stars \citep{Koenigl91,Matt_Pudritz05}. 
This problem, however, is highly non-trivial \citep[\emph{e.g.}][]
{Lubow_Spruit95,Heyvaerts_Priest_Bardou96,Livio_Ogilvie_Pringle99,
Spruit_Uzdensky05}, in part because the effective transport of the 
large-scale open flux may be greatly affected by reconnection with 
the small- and intermediate-scale closed coronal magnetic loops 
\citep{Spruit_Uzdensky05,Fisk05}. Incorporating a net large-scale 
flux into our statistical model is very straightforward and its overall 
transport should come out automatically.  We therefore believe that our 
theoretical model can be a very useful tool for addressing this problem.



\subsection{Twisted Loops}
\label{subsec-twisted}

In this paper we have assumed, for simplicity, that coronal loops have no 
longitudinal current, and hence have no internal twist. In a more general 
situation, however, there may be force-free field-aligned currents along 
the loops, generated in response to certain disk footpoint motions:
specifically, rotation of field-line footpoints around each other. 

Internal twist would have two major consequences for our coronal
model.  First, the pinch force of the associated longitudinal current
(parallel to the magnetic field because of the force-free assumption)
will tend to reduce the width of loop and its cross section for
reconnection with other loops.  One might consider as a limiting case
that the effective thickness of the loop is constant along its length.
There is indeed observational evidence that in the solar corona bright 
loops usually have a nearly constant thickness along their length 
\cite{Klimchuk00}.

Second, if the loop twist becomes too large, then the entire loop may
become kink-unstable and, as a result, the loop makes a transition to
a different equilibrium, where the internal twist is partially
transformed into the global writhe of the loop. 
That is, the loop no longer lies in one plane but rather has a twisted,
helical shape. Such $S$-like loops (so called sigmoidal loops) are
routinely observed in the solar corona \citep{Rust_Kumar96}.



\subsection{Incorporating Realistic Reconnection Physics
(Collisionless Reconnection Condition)}
\label{subsec-recn-collisionless}

The physics of reconnection is notoriously complex.
However, significant progress has been achieved in recent years and 
the picture that emerges can be summarized as follows~\citep{Uzdensky07b}. 
There are two regimes of reconnection: a slow (Sweet--Parker) regime 
in collisional plasmas and a fast regime in collisionless plasmas. 
The slow reconnection regime is just as important as the fast one, 
since, without it, it would be difficult for the system to accumulate  
significant free magnetic energy before releasing it suddenly {\it via} 
fast flare-like events. For practical purposes, the actual rate of fast 
reconnection is not very critical in our problem, as long as it is faster 
than the main dynamical time scale ({\it i.e.}, the orbital period). 
More important is the fast-reconnection {\it onset}, or {\it trigger}, 
problem, {\it i.e.}, the question of the transition from the slow to 
fast reconnection regime occur. 
The physics of fast collisionless reconnection is very complicated; 
it involves either two-fluid effects, such as the Hall effect, and/or 
anomalous resistivity due to current-driven plasma micro-instabilities.
However, despite this complexity, one can formulate a rough criterion 
for the transition from the slow collisional to the fast collisionless 
regime \citep{Cassak_Shay_Drake05,Yamada_etal06,Uzdensky07a,Uzdensky07b}. 
We thus plan to utilize this condition to formulate a physically-motivated 
prescription for handling reconnection. This prescription can then be used 
directly in our statistical theory or as a sub-grid model in actual MHD 
simulations of the corona. In our present model in this paper, the 
reconnection parameter~$\kappa$, which for simplicity we take to be 
constant, effectively subsumes all this complexity.

The condition for transition to fast collisionless reconnection 
involves several physical parameters of the system, including the 
ambient plasma density. The dependence of the reconnection regime 
on the density is critical, since it establishes an important feedback 
that the dynamically subdominant coronal gas exerts on the coronal 
magnetic field \citep{Uzdensky07a,Uzdensky07b}. 
In turn, the plasma density in the corona is determined by the disk--corona 
mass exchange processes, such as evaporation in response to coronal heating, 
precipitation due to gradual radiative cooling, and magneto-centrifugally 
and radiatively driven winds. Coupled together, disk--corona mass exchange 
and the transition to fast collisionless reconnection ensure that the corona
is maintained near the state of marginal collisionality and regulate the 
overall level of coronal activity and its intermittency, as well as the 
vertical distribution of magnetic energy density and of magnetic dissipation.

These ideas, especially the concept of marginal collisionality, have recently 
been successfully applied to the solar corona to explain the self-regulating 
nature of the coronal heating process \citep{Uzdensky07a,Uzdensky07b} and 
to the coronae of other main-sequence stars \citep{Cassak_Mullan_Shay07}. 
They also have proved very useful for providing a natural explanation for 
the observed optical depth in the coronae of accreting black holes (Goodman 
\& Uzdensky 2008, in preparation).


Finally, especially for studies of the ADC's interaction with 
a large-scale disk wind \citep[\emph{e.g.}][]{Brandenburg_vonRekowski07}, 
it is important to understand the transition from the force-free 
regime to the wind regime in which the plasma inertia becomes 
dynamically important \citep{Uzdensky_Koenigl_etal02b}. 
This transition happens near the Alfv\'en critical surface of 
the outflow and has a strong effect on magnetic reconnection. 
In particular, it is expected that, beyond the Alfv\'en surface, 
open magnetic field lines will not be able to close back {\it via} 
reconnection \cite[\emph{e.g.}][]{Uzdensky04}.




\subsection{Flux Emergence}
\label{subsec-flux-emergence}

Our model, by construction, is not complete---it needs to be connected
to what happens in the disk. In particular, it needs as input a
statistical description of the population of small loops, or of their
rate of emergence from the disk.  This information determines the overall 
normalization of the loop distribution function and hence the net rates of 
angular momentum transport and dissipation in the corona.  In the models
computed here, we have simply assumed a fixed, isotropic distribution
of small loops.

The ultimate source of coronal activity is the MHD turbulence in the disk 
\citep[\emph{e.g.}][]{Galeev_Rosner_Vaiana79,Tout_Pringle92,Miller_Stone00}. 
Therefore, in order to estimate the rate and form of magnetic flux 
emergence into the corona, one first needs to understand the properties 
of MHD turbulence in a stratified disk, with a particular emphasis on 
the production, evolution, and buoyant rise of magnetic flux tubes
\citep[\emph{e.g.}][]{Schramkowski_Achterberg93}.  The best prospect 
for developing this understanding is through appropriate statistical 
analyses of MRI turbulence in stratified shearing boxes.

However, the notion of a flux tube in most theoretical discussions 
(including the present paper) is not sufficiently precise to be applied 
directly to simulations.  
What well-defined, measurable, and statistically meaningful quantities
correspond to flux tubes or to their rates of emergence?  Candidates
exist, but it is not clear which is best.  Field lines can
be found as integral curves of the field and their motions determined,
but by what prescription should they be grouped into tubes?
Alternatively one might work with Fourier decompositions of the
vertical Poynting flux on horizontal planes, or with more general
$n$-point correlation functions of the field.


\section{Summary and Conclusions}
\label{sec-conclusions}

In this paper we construct a general theoretical framework for understanding 
the structure of a strongly-magnetized corona above a turbulent accretion 
disk. This study is motivated by the need to provide a more solid physical 
foundation for ADC spectral modeling efforts. It should also act as a 
connecting bridge between numerical MHD simulations of MRI-turbulent 
disks and semi-empirical coronal models, and stimulate further theoretical 
studies of accretion disks coupled to their coronae. We also hope that some 
of the theoretical tools and ideas developed in this paper will prove useful 
in solar physics.

One of the major goals of our study is to develop a statistical 
language appropriate for describing the chaotic coronal magnetic 
field. Here, we are interested in spatial scales larger than the 
disk thickness but smaller than its radius and in temporal scales 
longer than the orbital period but shorter than the overall accretion 
time. Our approach builds upon the previous work by Tout \& Pringle (1996),
but uses much more realistic physics in several key aspects and also 
goes further in analyzing and interpreting the results.

To construct the statistical theory, we represent the corona by an ensemble 
of elementary magnetic structures, namely, loops connecting two spots on 
the disk surface (Fig.~\ref{fig-corona}). Each loop is characterized by 
several primary parameters ({\it e.g.}, the distance between the footpoints
and the orientation). The main object in this study is the {\it distribution 
function}~$F$ of loops in this parameter space. One of our main goals is to 
formulate, and then solve, the {\it loop kinetic equation} (LKE) for this 
distribution function, similar to the Boltzmann kinetic equation in the 
statistical theory of gases. 

To do this, we first analyze the key physical processes that govern 
the evolution of individual coronal loops. First, there are several 
processes that affect the loops individually, such as:
(1) emergence of small loops into corona;
(2) random footpoint motions due to the disk turbulence;
(3) Keplerian shear, stretching loops azimuthally and thereby 
also making them grow in height.
On average, these processes pump energy from the disk into the corona, 
creating a stressed non-potential force-free field. In addition, loops 
may interact with each other {\it via} episodic reconnection between two 
individual loops \citep[\emph{c.f.}][]{Tout_Pringle96}, forming two new loops 
(see Fig.~\ref{fig-recn}). In our theory reconnection is represented 
as a binary collision, analogous to binary collisions between atoms 
in a gas. Correspondingly, we describe loop-loop reconnection by a 
nonlinear integral operator, similar to Boltzmann's collision operator.
In contrast to processes (1)--(3), magnetic reconnection relaxes 
the accumulated magnetic stresses and dissipates the accumulated 
free magnetic energy. Overall, a magnetically-active ADC can be 
described as a Boiling Magnetic Foam.

Based on these processes we are able to construct the loop kinetic 
equation. In this equation we characterize the overall rate of 
reconnection events relative to the Keplerian shear rate by a 
dimensionless parameter~$\kappa$. In order to investigate the 
role of magnetic reconnection in the corona, we solve the loop
kinetic equation numerically for several different values of~$\kappa$.
We obtain a {\it statistical steady state} for each value of~$\kappa$
and find that the steady-state loop distribution function is generally 
well represented by a orientation-dependent power-law, $F(L,\theta) 
\sim L^{-\alpha_\kappa(\theta)}$. When Keplerian shear is absent,
the distribution function is isotropic, $\alpha_\infty(\theta)=
{\rm const}$. As the rate of shear relative to reconnection increases
({\it i.e.}, $\kappa$ decreases), the distribution becomes more and more
anisotropic, with a predominance of toroidal orientation. At the same
time, a typical loop grows to a larger size by a stronger shear before 
its growth is disrupted by reconnection. Thus, the orientation-averaged
distribution function becomes shallower as~$\kappa$ is decreased.

Once the distribution function is known, we use it to calculate several
important integral quantities related to the energetics of the corona.
First, we use a self-consistent {\it mean-field} approach to compute 
the magnetic energy density as a function of height, $\bar{B}^2(z)/8\pi$. 
This quantity represents the collective magnetic pressure of all the 
neighboring loops that confine any given loop and thus represents another 
(in addition to reconnection) way in which loops interact with each other 
in our theory. Although it doesn't enter explicitly into the loop kinetic 
equation, $\bar{B}(z)$ is very important in our model. In particular, it 
controls the thickness of loops as a function of height, which affects in 
turn the cross-section for reconnection. In addition, by requiring that
the vertical gradient of the magnetic pressure $\bar{B}^2(z)/8\pi$ be
balanced by the magnetic tension within each loop, we self-consistently
calculate the equilibrium shape and vertical extent, $Z_{\rm top}(L)$, 
of the loops. This, in turn, enables us to calculate some important 
quantities such as the energy associated with a given loop, the force 
exerted on its footpoints, etc. We then use these quantities to assess 
various issues of coronal energetics, including the overall magnetic 
energy stored in the corona, statistical distribution of coronal energy 
release events (flares), and the overall angular momentum transferred 
by the coronal magnetic field. As a result of our parametric study with 
respect to the reconnection parameter~$\kappa$, we find that if~$\kappa$ 
is decreased ({\it i.e.}, reconnection in the corona is inhibited) beyond 
a certain value, the slope of the loop distribution function becomes so
shallow (namely, shallower than~$L^{-3/2}$) that the contribution of large 
loops to both the magnetic energy and torque starts to dominate, leading 
to a significant enhancement in these quantities. In our specific model, 
the critical value of~$\kappa$ is found to be $\kappa_{3/2}\simeq 0.002$.

These results demonstrate that the energetic dominance of coronae
is inextricably linked to reconnection processes. 
They thus motivate further efforts to develop more realistic physical 
description of reconnection. 
To reiterate an important point made in the Introduction, the tenuous 
corona above an accretion disk is likely to be marginally collisionless 
(Goodman \& Uzdensky 2008, in preparation), unlike the dense plasma inside 
the disk itself. This means that the corona cannot be described by 
traditional MHD simulations with constant resistivity because of their 
inability to control or resolve magnetic reconnection, which, as we have 
shown in this paper, may influence the coronal magnetic energy and angular 
momentum transfer. Therefore, some kind of a physically-motivated subgrid 
prescription for reconnection is needed.


\begin{acknowledgments}

We would like to acknowledge fruitful discussions 
with V.~Titov, Z.~Mikic, A.~Pankin, and D.~Schnack.
We are grateful to R.~Blandford, S.~Cowley, and V.~Titov 
for drawing our attention to several useful references.

This work is supported by National Science Foundation Grant 
No.\, PHY-0215581 (PFC: Center for Magnetic Self-Organization 
in Laboratory and Astrophysical Plasmas).

\end{acknowledgments}

\bibliographystyle{apj}
\bibliography{xxx2}


\appendix
\section{Appendix A: Calculation of $\bar{B}$ in terms of~$\bar{F}(L)$}
\label{appendix-A}

In this Appendix we calculate the mean magnetic field $\bar{B}(L)$ 
in terms of the loop distribution function $\bar{F}(L)$ corresponding 
to a self-consistent atmosphere (see \S~\ref{subsec-Bbar}).

\begin{figure}
\plotone{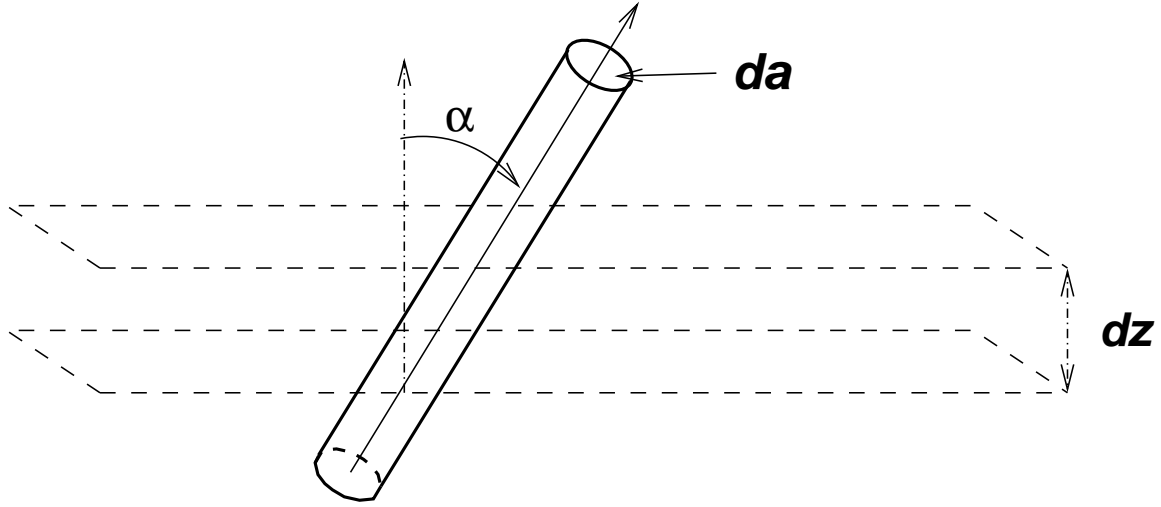}
\figcaption{A loop segment of area $a(z)$ and angle~$\alpha$, 
crossing a horizontal slab of thickness~$dz$.
\label{fig-slab}}
\end{figure}

Consider a horizontal slab of thickness~$dz$ at height~$z$ 
(see Fig.~\ref{fig-slab}). Correspondingly, there is a minimal 
length~$L$ of loops that reach above this height. Consider now 
an arbitrary slender loop~$\mathcal{A}$ of length $L'>L$, with 
a cross-sectional area~$a(z)$ and the angle between the loop's 
direction and the vertical equal to~$\alpha(z,L')$. Then the 
volume that the two segments (ascending and descending) of this 
loop occupy inside the slab under consideration can be written as
\beq
dV = 2 a(z)\  {dz\over{\cos{\alpha(z,L')}}} \, .
\eeq
The volume occupied by all such loops per unit horizontal area 
of the slab is equal to
\beq
dV =  2 \int \limits_L^\infty \, dL'
\ a(z) \, \bar{F}(L')\ {dz\over{\cos{\alpha(z,L')}}} \, .
\eeq
Since in our model each loop carries the same amount~$\Delta\Psi$ 
of magnetic flux, the loop cross-sectional area $a(z)$ is the same 
for all loops at a given height and is simply equal to 
$\Delta\Psi/\bar{B}(z)$. Thus, the above volume is
\beq
dV =  {2\Delta\Psi\over{\bar{B}(z)}}\  dz \int \limits_L^\infty 
\bar{F}(L')\ {{dL'}\over{\cos{\alpha(z,L')}}} \, .
\eeq

But, on the other hand, these loops occupy all the volume within the slab, 
and so this volume per unit horizontal area ought to be equal to just~$dz$. 
This gives us the equation that determines $\bar{B}$ as a function of~$L$:
\beq
\bar{B}(z) = B_0 b(z) = 2\Delta\Psi \int \limits_L^\infty 
\bar{F}(L') \ {{dL'}\over{\sqrt{1-\sin^2{\alpha(z,L')}}}} \, .
\eeq

Now, using equation~(\ref{eq-loopshape-2}), we can replace
$\sin{\alpha(z,L')} = b_{\rm top}(L')/b(z) = b_{\rm top}(L')/b_{\rm top}(L)$
and hence obtain the following integral equation for the function~$b(L')$:
\beq
\int \limits_L^\infty 
{{\bar{F}(L')\,dL'}\over{\sqrt{b_{\rm top}^2(L)-b_{\rm top}^2(L')}}} = 
{{B_0}\over{2\Delta\Psi}} = {\rm const} \, ,
\eeq
which can be rewritten as 
\beq
\int \limits_0^b 
{{U(b')\, db'}\over{\sqrt{b^2-b'^2}}} = 
{{B_0}\over{2\Delta\Psi}} = {\rm const}
\eeq
where $U(b')\equiv - \bar{F}[L(b')]\, dL/db'$. By substitutions $t=b'^2$,
$s=b^2$, and $V(t)=U(b')/2b'$, this equation can be transformed into the 
Abel integral equation, that can be immediately solved, yielding the 
following final result:
\beq
V(s) = {{B_0}\over{2\pi\Delta\Psi \sqrt{s}}} 
\quad \Rightarrow \quad 
U(b)= -\, \bar{F}[L(b)] \, dL/db  = {B_0\over{\pi\Psi_0}} \, .
\eeq
and hence
\beq
db = -\, {{\pi\,\Psi_0}\over{B_0}}\ \bar{F}(L)\, dL \quad  \Rightarrow \quad 
b(L) =  {{\pi\,\Psi_0}\over{B_0}} \ 
\int\limits_L^\infty \bar{F}(L') \, dL' \, , 
\eeq
which is in agreement with our general expectation above.


\section{Appendix B: Proof that $\mathcal{E}(L) = 2 E_{\rm magn}$}
\label{Appendix-B}

In this Appendix we prove the conjecture that 
$\mathcal{E}(L) = 2 E_{\rm magn}(L)$. The proof 
goes as follows.

First, according to equation~(\ref{eq-f_footpt}), the force on 
the loop's footpoint can be expressed in terms of the magnetic 
field at the top of the loop as
\beq
f_{\rm fp}(L) = {{\Delta\Psi B_{\rm hor}(z=0;L)}\over{4\pi}} = 
{{\Delta\Psi B_{\rm top}(L)}\over{4\pi}} \, .
\label{eq-force-fp}
\eeq

Using expression~(\ref{eq-E_magn-2}) and taking into
account that a loop has two legs, we have
\beq
E_{\rm magn}(L) = 
2\times {\Delta\Psi\over{8\pi}}\, \int\limits_{\rm left\ leg} B(l)\, dl =
{\Delta\Psi\over{4\pi}}\, \int\limits_{z=0}^{z_{\rm top}(L)} \ 
\bar{B}(z)\, {dz\over{\cos\alpha}} = 
{\Delta\Psi\over{4\pi}}\, \int\limits_{z=0}^{z_{\rm top}(L)} \ 
{{\bar{B}^2(z)}\over{\sqrt{\bar{B}^2-B_{\rm top}^2}}}\ dz \, .
\label{eq-E_magn-3}
\eeq

On the other hand, substituting~(\ref{eq-force-fp}) 
into equation~(\ref{eq-energy-1}), we get
\beq
\mathcal{E}(L) =
{{\Delta\Psi B_0}\over{4\pi}}\ \int\limits_0^L\, b_{\rm top}(L')\, dL' \, . 
\eeq

Integrating this by parts yields
\beq 
\mathcal{E}(L) = \int\limits_0^L b(L')\, dL' =
b(L)\, L + \int\limits_{b(L)}^1 L(b')\, db' \, .
\eeq

According to (\ref{eq-z_top-1}):
\beq
\int\limits_{b(L)}^1 L(b')\, db' = 
-\, 2\ \int \limits_{b'=b(L)}^1 b' \int \limits_{b'}^1 \ 
{{[dz(b'')/db'']}\over{\sqrt{b''^2 - b'^2}}}\ db''\, db' \, .
\eeq

Interchanging the order of integration, we get
\begin{eqnarray}
\int\limits_{b}^1 L(b')\, db' &=& 
-\, 2 \int \limits_{b''=b}^1\, db'' \ {{dz(b'')}\over{db''}}\ 
\biggl(\int \limits_{b'=b}^{b''} {b'\over{\sqrt{b''^2-b'^2}}}\, db' \biggr) 
\nonumber \\ 
&=& -\, 2 \int \limits_{b''=b}^1\, 
\sqrt{b''^2 - b^2} \ {{dz(b'')}\over{db''}}\ db''\, .
\end{eqnarray}

On the other hand, according to (\ref{eq-z_top-1}),
\beq
b(L) \, L =  -\, 2 \int \limits_{b''=b(L)}^1\   
{{b^2\, [dz(b'')/db'']}\over{\sqrt{b''^2 - b^2}}} \,  db'' \, .
\eeq

Combining these results we get
\beq 
\mathcal{E}(L) = -\, 2 \int \limits_{b''=b(L)}^1\   
{{b''^2\, dz''}\over{\sqrt{b''^2 - b^2}}}  = 2 \, E_{\rm magn} \, .
\eeq

End of Proof.

\end{document}